\documentclass[12pt]{elsarticle}
 \usepackage{ifpdf,amssymb,amsmath,epsfig,bm,pifont}
  \graphicspath{{./figs/}}
   \pdfoutput=1
\usepackage{color}

\renewcommand{\thefootnote}{\fnsymbol{footnote}}

\biboptions{compress}

\begin{document}
\begin{frontmatter}

\title{\vskip-3cm{\baselineskip14pt
\centerline{\normalsize\hfill arXiv:1204.2679}
}\vskip.7cm
\texttt{FAPT}: a \texttt{Mathematica} package for calculations
in QCD Fractional Analytic Perturbation Theory
\vskip.3cm
}
 \author[jinr]{Alexander~P.~Bakulev}
  \ead{bakulev@theor.jinr.ru}

 \author[jinr,gstu]{Vyacheslav~L.~Khandramai}
  \ead{v.khandramai@gmail.com}

\address[jinr]{Bogoliubov Laboratory of Theoretical Physics, JINR,
                141980 Dubna, Russia\\}
\address[gstu]{Gomel State Technical University,
               246746 Gomel, Belarus\\}
\date{}

\begin{abstract}
\noindent
We provide here all the procedures in \texttt{Mathematica}
which are needed for the computation
of the analytic images of the strong coupling constant powers
in Minkowski
(${\bar{\mathfrak A}_{\nu}(s;n_f)}$ and
 ${\mathfrak A_{\nu}^\text{glob}(s)}$)
and Euclidean
(${\bar{\mathcal A}_{\nu}(Q^2;n_f)}$ and
 ${\mathcal A_{\nu}^\text{glob}(Q^2)}$)
domains
at arbitrary energy scales
(${s}$ and ${Q^2}$, correspondingly)
for both schemes --- with fixed number of active flavours ${n_f=3, 4, 5, 6}$
and the global one with taking into account all heavy-quark thresholds.
These singularity-free couplings
are inevitable elements
of Analytic Perturbation Theory (APT) in QCD~\cite{SS,MS96,SS98},
and its generalization ---
Fractional APT~\cite{BMS05,BKS05,BMS06},
needed to apply the APT imperative for renormalization-group improved
hadronic observables.
\vspace{.2cm}


\end{abstract}

\end{frontmatter}

\thispagestyle{empty}
 \newpage
  \setcounter{page}{1}

\renewcommand{\thefootnote}{\arabic{footnote}}
 \setcounter{footnote}{0}

\section*{Program Summary}
\begin{itemize}

\item[]\textit{Title of program:}
  \texttt{FAPT}

\item[]\textit{Available from:}\\
  \texttt{http://theor.jinr.ru/$\tilde{\phantom{x}}$bakulev/fapt.mat/FAPT.m
  }\\
  \texttt{http://theor.jinr.ru/$\tilde{\phantom{x}}$bakulev/fapt.mat/FAPT\_Interp.m
  }

\item[]\textit{Computer for which the program is designed and others on which it
    is operable:}
  Any work-station or PC where \texttt{Mathematica} is running.

\item[]\textit{Operating system or monitor under which the program has been
    tested:}
  Windows XP, Mathematica (versions 5 and 7).

\item[]\textit{No. of bytes in distributed program including test data etc.:}\\
  $47$\,kB (main module \texttt{FAPT.m}) and $4$\,kB (interpolation module \texttt{FAPT\_Interp.m});\\
  $21$\,kB (notebook \texttt{FAPT\_Interp.nb} showing how to use the interpolation module);\\
  $10\,888$\ kB (interpolation data files: \texttt{AcalGlob}$\ell$\texttt{i.dat}
                  and \texttt{UcalGlob}$\ell$\texttt{i.dat}
                  with $\ell=1$, $2$, $3$, $3$\text{P}, and $4$)\footnote{
                  The notebook \texttt{FAPT\_Interp.nb} and all interpolation data files are
                  available from the same place in the form of the zipped archive \texttt{FAPT\_Interp.zip}
                  of the size 1844\,kB.
                  In order that \texttt{Mathematica} notebook \texttt{FAPT\_Interp.nb}
                  can use these precalculated data files one should place the directory
                  \texttt{$.\backslash$sources$\backslash$}
                  with all data files in the same directory as the main file \texttt{FAPT\_Interp.nb}.}

\item[]\textit{Distribution format:}
  ASCII

\item[]\textit{Nature of physical problem:}
  The values of analytic images ${\bar{\mathcal A}_{\nu}(Q^2)}$
  and ${\bar{\mathfrak A}_{\nu}(s)}$
  of the QCD running coupling powers ${\alpha_\text{s}^{\nu}(Q^2)}$
  in Euclidean and Minkowski regions,
  correspondingly,
  are determined through the spectral representation
  in the QCD Analytic Perturbation Theory (APT).
  In the program \texttt{FAPT} we collect all relevant formulas
  and various procedures
  which allow for a convenient evaluation of
  ${\bar{\mathcal A}_{\nu}(Q^2)}$
  and ${\bar{\mathfrak A}_{\nu}(s)}$
  using numerical integrations of the relevant spectral densities.

\item[]\textit{Method of solution:}
  \texttt{FAPT} uses \texttt{Mathematica} functions to calculate
  different spectral densities and then performs
  numerical integration of these spectral integrals to obtain
  analytic images of different objects.

\item[]\textit{Restrictions on the complexity of the problem:}
  It could be that for an unphysical choice of the input parameters
  the results are out of any meaning.

\item[]\textit{Typical running time:}
  For all operations the running time does not exceed a few seconds.
  Usually numerical integration is not fast,
  so that we advice to use arrays of precalculated data
  and apply then the routine \texttt{Interpolate}
  (as shown in supplied example of the program usage,
  namely in the notebook \texttt{FAPT\_Interp.nb}).
\end{itemize}

\section{Introduction}
 \label{sec:intro}

QCD perturbation theory (PT) in the region of spacelike four-momentum transfer
($Q^2 = -q^2 > 0$ --- hereafter we call it the Euclidean region)
is based on expansions in a series over the powers of effective charge
(or running coupling constant) $\alpha_s(Q^2)$,
which in the one-loop approximation
is given by $\alpha_s^{(1)}(Q^2) = (4\pi/b_0)/L$
with $b_0$ being the first coefficient of the QCD beta function,
Eq.\,(\ref{eq:beta})--(\ref{eq:beta.coef}),
$L = \ln(Q^2/\Lambda^2)$,
and $\Lambda = \Lambda_\text{QCD}$ is the QCD scale parameter.
The one-loop solution $\alpha_s^{(1)}(Q^2)$ has a pole singularity
at $L=0$ called the Landau pole.
The $\ell$-loop solution $\alpha_s^{(\ell)}(Q^2)$
of the renormalization group equation (\ref{eq:beta})
has an $\ell$-root singularity of the type $L^{-1/\ell}$
at $L=0$,
which produces the pole as well in the $\ell$-order term
$d_\ell\,\alpha_s^\ell(Q^2)$.
This prevents the application of perturbative QCD
in the low-momentum spacelike regime,
$Q^2\sim\Lambda^2$,
with the effect that hadronic quantities,
calculated at the partonic level
in terms of a power-series expansion in the running coupling,
are not everywhere well defined.

Such a singularity appeared first in QED~\cite{LAH56,DG05}
and was named ``ghost'' due to the negative residue
at the corresponding propagator pole.
It was interpreted as an indication
that quantum field theory is self-contradictory.
However, as was shown in~\cite{BS59,BSvvtkp},
it is only a hint about the PT inapplicability
in the region
where the expansion parameter is not small.
Appearance of such ``ghost'' singularities
from a theoretical point of view
contradicts the causality principle in quantum field theory~\cite{BSvvtkp,BLT69},
since it makes the K\"{a}llen--Lehmann spectral representation
impossible.
It also complicates the determination
of the effective charge in the timelike region
($q^2>0$ --- hereafter we call it the Minkowski region).
In a seminal paper by N.~N.~Bogoliubov et al. of 1959~\cite{BLS60},
the ghost-free effective coupling for QED has been constructed
using the dispersion relation technique.

After the very appearance of QCD
many researchers tried to determine
the QCD effective charge in the Minkowski region,
which is suitable for describing the processes
of $e^+e^-$ annihilation into hadrons,
as well as quarkonium and $\tau$-lepton decays into hadrons.
Many such attempts used analytic continuation of the effective charge
from the deep Euclidean region,
in which perturbative QCD is known to work well,
into a Minkowski one,
where actual experiments were performed:
$\alpha_s(Q^2)\to\alpha_s(s=-Q^2)$.
In 1982 Radyushkin \cite{Rad82}
and Krasnikov and Pivovarov \cite{KP82}
using the dispersion technique of~\cite{BLS60}
suggested regular (for $s\geq \Lambda^2$)
QCD running coupling in Minkowski region,
the well-known $\pi^{-1}\arctan(\pi/L)$.

In 1995 Jones and Solovtsov using variational approach~\cite{JS95-349}
constructed the effective couplings in Euclidean and Minkowski domains
which appears to be finite for all $Q^2$ and $s$
and satisfy analyticity integral conditions.
Just in the same time Shirkov and Solovtsov~\cite{SS},
using the dispersion approach of~\cite{BLS60},
discovered ghost-free coupling $\mathcal A_1(Q^2)$, Eq.\ (\ref{eq:A_1}),
in Euclidean region
and
ghost-free coupling $\mathfrak A_1(s)$, Eq.\ (\ref{eq:U_1}),
in Minkowski region,
which satisfy analyticity integral conditions:
\begin{eqnarray}
\label{eq:R-D-operation}
 \mathcal A_1(Q^2)
  = Q^2~\int_0^{\infty}
    \frac{\mathfrak A_1(\sigma)}{(\sigma+Q^2)^2}
     d\sigma\,;\quad\quad
 \mathfrak A_1(s)
  = \frac{1}{2\pi i}
     \int_{-s-i\varepsilon}^{-s+i\varepsilon}\!
      \frac{\mathcal A_1(\sigma)}{\sigma}\,
  d\sigma\,.
\end{eqnarray}
At the one-loop approximation
the last coupling coincides with the Radyushkin one
for $s\geq\Lambda^2$.
This way of making the QCD's effective charge
analytic in the timelike region
was rediscovered later
within an approach of fermion bubble resummation
by Beneke and Braun~\cite{BB95},
and also
by Ball, Beneke, and Braun~\cite{BBB95}.
Due to the absence of singularities in these couplings,
Shirkov and Solovtsov suggested
to use this systematic approach,
called Analytic Perturbation Theory (APT),
for all $Q^2$ and $s$.

Recently the analytic and numerical methods,
necessary to perform calculations in two- and three-loop approximations,
were developed~\cite{Mag99,Kour99,Mag00,KM01,Mag03u,KM03,Mag05}.
This approach was applied to the calculation of properties
of a number of hadronic processes,
including the width of inclusive $\tau$ lepton decay to hadrons~\cite{MSSY00,MSS01,CvVa06,CKV09,Mag10},
the scheme and renormalization-scale dependencies
in the Bjorken~\cite{MSS98,PST08}
and Gross--Llewellyn Smith~\cite{MSS98GLS} sum rules,
the width of $\Upsilon$ meson decay to hadrons~\cite{SZ05},
etc.
Moreover, APT was applied to the analysis of the processes
with two scales rather than just a single scale, namely:
the pion-photon transition form factor~\cite{SSK99,SSK00}
and the pion electromagnetic form factor
in the $O(\alpha_s)$ order~\cite{SSK99,SSK00,BPSS04}.
To summarize,  we can say that APT
(see reviews~\cite{SS99,Shi00,SS06})
yields a sensible description of hadronic quantities in QCD,
though there are alternative approaches
to the singularity of effective charge
in QCD --- in particular,
with respect to the deep infrared region $Q^2<\Lambda^2$,
where appearance of nonzero hadronic masses
may be important~\cite{DPTar89,Sim01,NP04}.
The main advantage of the APT analysis
is much more faster convergence
of the APT non-power series
as compared with the standard PT power series,
see in~\cite{BaSh11,Shi12}.

Three-point functions,
used in describing the pion electromagnetic form factor
or $\gamma^*\gamma\to\pi^0$ transition form factor,
contain logarithmic contributions
at the next-to-leading order
of the QCD PT,
related to the factorization scale.
If one set the factorization scale proportional
to the squared momentum-transfer,
$\mu_\text{F}^2=Q^2$,
then these logarithms will go to zero,
but additional RG factors
of the type $(\alpha_s(Q^2)/\alpha_s(\mu_0^2))^\nu$,
with $\nu=\gamma_n/(2_0)$ being a fractional number,
will appear
in the Gegenbauer coefficients of the pion distribution amplitude.
In both cases
spectral densities,
used to construct analytic images of hadronic amplitudes,
should change.
This observation led Karanikas and Stefanis~\cite{KS01,Ste02}
to propose the concept of analytization ``as a whole'',
meaning that one should construct analytic images
not only of effective charge and its powers,
but of the whole QCD amplitude under consideration.

A QCD inspired generalization of APT to the fractional powers of effective charge,
called Fractional Analytic Perturbation Theory (FAPT),
was done in~\cite{BMS05,BMS06}
(for a recent review see~\cite{AB08},
for a recent generalization see~\cite{CvKo11fapt}),
followed by the application~\cite{BKS05}
to the analysis of the factorizable contribution
to the pion electromagnetic form factor.
The crucial advantage of FAPT
in this case
is that the perturbative results start
to be less dependent on the factorization scale.
This reminds the results,
obtained with the APT,
applied to the analysis of the pion form factor in the $O(\alpha_s^2)$ approximation,
where the results also almost cease to depend
on the choice of the renormalization scheme and its scale
(for a detailed review see~\cite{AB08} and references therein).
The process of the Higgs boson decay
into a $b\bar{b}$ pair of quarks
was studied within a framework of FAPT
in the Minkowski region
at the one-loop level in~\cite{BKM01}
and at the three-loop level--- in~\cite{BMS06}.
Results on the resummation of non-power-series expansions of the Adler function
of a scalar, $D_S$, and a vector, $D_V$, correlators within FAPT
were presented in~\cite{BMS10}.
The interplay between higher orders of the perturbative QCD (pQCD) expansion
and higher-twist contributions in the analysis of recent Jefferson Lab
data on the lowest moment of the spin-dependent proton structure function,
$\Gamma_1^{p}(Q^2)$,
was studied in~\cite{PSTSK09} using both standard QCD PT and (F)APT.
FAPT technique was also applied to the analysis
of the structure function $F_2(x)$ behavior at small values of $x$~\cite{CIKK09,KotKri10}.
All these successful applications of (F)APT
necessitate to have a reliable mathematical tool
for calculations of spectral densities and analytic couplings
which are implemented in FAPT.\footnote{
This task has been partially realized for both APT and its massive generalization~\cite{NP04}
as the \texttt{Maple} package \texttt{QCDMAPT} in~\cite{NeSi10} and
as the \texttt{Fortran} package \texttt{QCDMAPT\_F} in~\cite{NeSi11}.
Both these realizations are limited to the case of fixed number of active quarks $N_f=3$ only,
and use approximate expressions for the two- and higher-loop perturbative couplings,
compare, for example, Eq.\,(33) in \cite{NeSi10} and our Eq.\,(\ref{eq:a.2L}).}

In this paper we collect all relevant formulas
which are necessary
for the running of $\bar{\mathcal A}_{\nu}[L]$ and $\bar{\mathfrak A}_{\nu}[L]$
in the framework
of APT and its fractional generalization, FAPT.
We discuss their proper usage
and provide easy-to-use \texttt{Mathematica}~\cite{math88}
procedures
collected in the package \texttt{FAPT}.
A few examples are given.
Here we do not consider the inclusion of analytic images
of logarithms multiplied by fractional powers of couplings,
namely,
$\left[\alpha_s(Q^2)\right]^{\nu}\cdot\left[\ln(Q^2/\Lambda^2)\right]^{m}$,
which are needed for the full implementation of FAPT, ---
we postpone it to the next paper.

The outline of the paper is as follows.
In the next Section we present
the main formulas of perturbative QCD
which are needed
for the running of the strong coupling constant
up to the four-loop level.
Section~\ref{sec:FAPT} contains
the basic formulas of APT and FAPT.\footnote{
Note here that FAPT includes APT as a partial case
for the integer values of indices.}
Finally, in Section~\ref{sec:proc}
we describe the most important procedures
of the package \texttt{FAPT}
and provide an example of using this package
to produce some numerical estimations.
We hope that for most practical applications it should be sufficient.
In the Appendix
we supply
the complete collection of the developed procedures.

\section{Basics of the QCD running coupling}
 \label{sec:pQCD}
The running of the coupling constant of QCD,
$\alpha_\text{s}(\mu^2)=\alpha_\text{s}[L]$
with
$L=\ln[\mu^2/\Lambda^2]$,
is defined through\footnote{
We use notations $f(Q^2)$ and $f[L]$ in order to specify the arguments we mean ---
squared momentum $Q^2$ or its logarithm $L=\ln(Q^2/\Lambda^2)$,
that is $f[L]=f(\Lambda^2\cdot e^L)$ and $\Lambda^2$ is usually referred to $n_f=3$ region.}
\begin{eqnarray}
 \label{eq:beta}
  \frac{d \alpha_\text{s}[L]}{d L}
    &=&
    \beta\left(\alpha_\text{s}[L];n_f\right)
    \,\,=\,\,
   -\,\alpha_\text{s}[L]\,
     \sum_{k\ge0}b_k(n_f)\,
      \left(\frac{\alpha_\text{s}[L]}{4\pi}\right)^{k+1}\,,
\end{eqnarray}
where $n_f$ is the number of active flavours.
The coefficients are given by
\cite{GW73,GW73a,Pol73,Jon74,Cas74,EgTar78,TVZ80,LaVe93,vRVL97,Cz04}
\begin{eqnarray}
 b_0(n_f)
  &=& 11 - \frac{2}{3} n_f\,,
  \nonumber\\
 b_1(n_f)
  &=& 102 - \frac{38}{3} n_f\,,
  \nonumber \\
 b_2(n_f)
  &=& \frac{2857}{2}
    - \frac{5033}{18} n_f
    + \frac{325}{54} n_f^2\,,
  \nonumber \\
 b_3(n_f)
  &=& \frac{149753}{6}
    + 3564\,\zeta_3
    - \left[\frac{1078361}{162} + \frac{6508}{27}\,\zeta_3\right] n_f
  \nonumber\\
  & &
  \label{eq:beta.coef}
    + \left[\frac{50065}{162} + \frac{6472}{81}\,\zeta_3 \right] n_f^2
    + \frac{1093}{729}  n_f^3\,.
\end{eqnarray}
$\zeta$ is Riemann's zeta function, with values $\zeta_2=\pi^2/6$
and $\zeta_3\approx1.202\,057$.
It is convenient to introduce the following notations:
\begin{eqnarray}
 \label{eq:ck.def}
  \beta_f\equiv \frac{b_0(n_f)}{4\pi}\,,\quad
  a(\mu^2;n_f)
   \equiv \beta_f\,\alpha_\text{s}(\mu^2;n_f)
   \quad\text{and}\quad
  c_k(n_f)
   \equiv \frac{b_k(n_f)}{b_0(n_f)^{k+1}}\,.
\end{eqnarray}
Then Eq.\,(\ref{eq:beta}) in the $l$-loop approximation
can be rewritten in
the following form:
\begin{eqnarray}
 \label{eq:beta.norm}
  \frac{d a_{(\ell)}[L;n_f]}{d L}
    &=&
    -\,\left(a_{(\ell)}[L;n_f]\right)^{2}\,
     \left[1+\sum_{k\ge1}^{\ell}c_k(n_f)\,\left(a_{(\ell)}[L;n_f]\right)^{k}\right]\,.
\end{eqnarray}
In the one-loop ($l=1$) approximation ($c_k(n_f)=b_k(n_f)=0$ for all $k\geq1$)
we have a solution
\begin{eqnarray}
 \label{eq:a.1L}
  a_{(1)}[L]
   &=& \frac{1}{L}
\end{eqnarray}
with the Landau pole singularity at $L\to0$.
In the two-loop ($l=2$) approximation ($c_k(n_f)=b_k(n_f)=0$ for all $k\geq2$)
the exact solution of Eq.\,(\ref{eq:beta})
is also known~\cite{Mag98,GGK98}
\begin{eqnarray}
 \label{eq:a.2L}
  a_{(2)}[L;n_f]
   = \frac{-c_1^{-1}(n_f)}
          {1 + W_{-1}\left(z_W[L]\right)}
   \quad\text{with}\quad
  z_W[L]
  = -c_1^{-1}(n_f)\,e^{-1-L/c_1(n_f)}
     \,,
\end{eqnarray}
where $W_{-1}[z]$ is the appropriate branch of Lambert function.

The three- and higher-loop solutions $a_{(\ell)}[L;n_f]$ can be expanded
in powers of the two-loop one, $a_{(2)}[L;n_f]$,
as has been suggested in~\cite{Kour99,Mag03u,KM03,Mag05,Mag10}:
\begin{equation}
 \label{eq:a.lL_in_2L}
  a_{(\ell)}[L;n_f]
   = \sum_{n\geq1} C_{n}^{(\ell)}\,\left(a_{(2)}[L;n_f]\right)^n.
\end{equation}
Coefficients $C_{n}^{(\ell)}$ are known
and can be evaluated recursively.
We use in our routine for the three-loop coupling
expansion
up to the 9-th power
included:
\begin{eqnarray}
 \label{eq:c(3)n}
  C_{1}^{(3)} = 1\,,\quad
  C_{2}^{(3)} = 0\,,\quad
  C_{3}^{(3)} = c_2\,,\quad
  C_{4}^{(3)} = 0\,,\quad
  C_{5}^{(3)} = \frac{ 5}{ 3}\,c_2^2\,,\quad
  C_{6}^{(3)} = \frac{-1}{12}\,c_1\,c_2^2\,,~~~
  \nonumber\\
  C_{7}^{(3)} = \frac{ 1}{20}\,c_1^2\,c_2^2
              + \frac{16}{ 5}\,c_2^3\,,\quad
  C_{8}^{(3)} = \frac{-1}{30}\,c_1^3\,c_2^2
              - \frac{23}{60}\,c_1\,c_2^3\,,
  ~~~~~~~~~~~~~~~~~~~~~~
  \nonumber\\
  C_{9}^{(3)} = \frac{ 1}{42}\,c_1^4\,c_2^2
              + \frac{103}{420}\,c_1^2\,c_2^3
              + \frac{2069}{315}\,c_2^4\,.
  ~~~~~~~~~~~~~~~~~~~~~~~~~~~~~~~~
\end{eqnarray}
\begin{figure}[t!]
 \centerline{\includegraphics[width=0.45\textwidth]{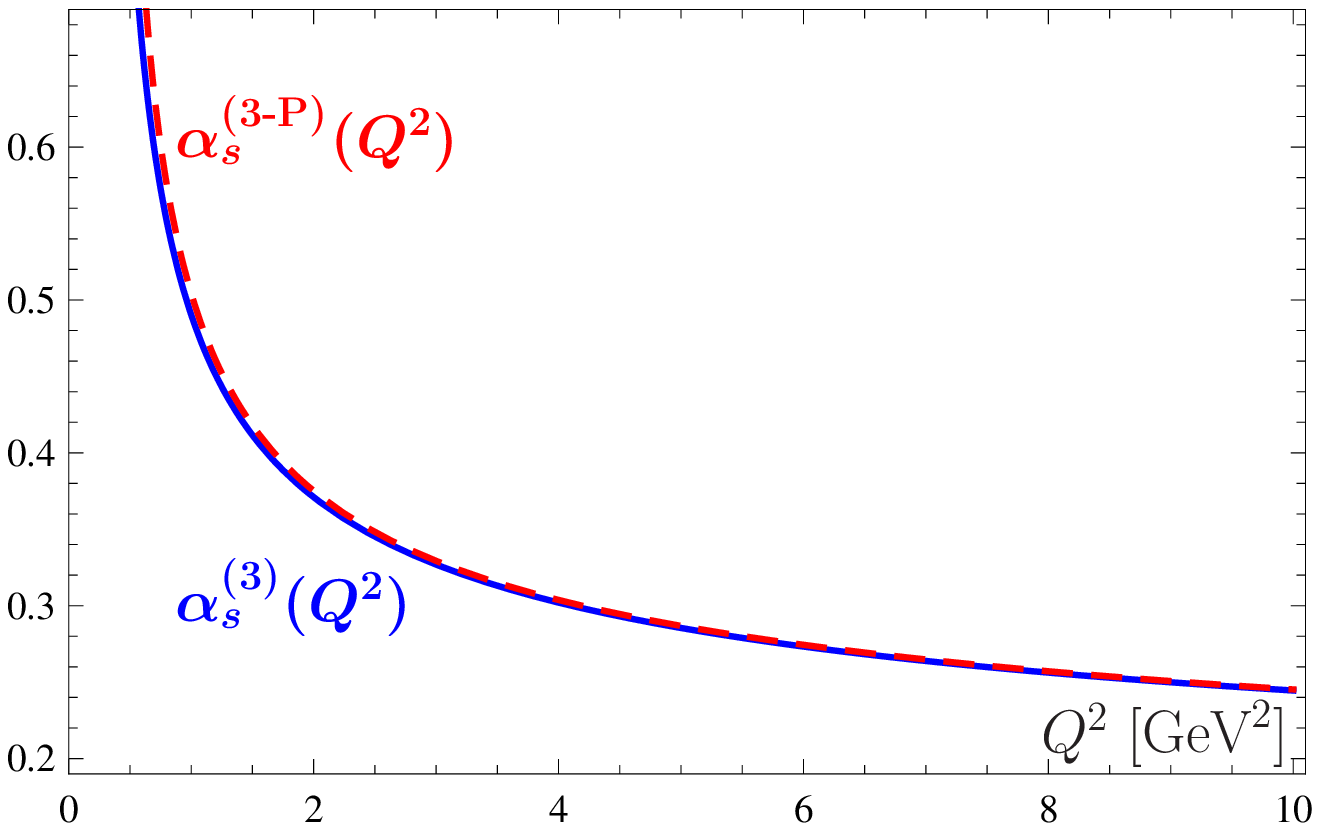}~~~
             \includegraphics[width=0.45\textwidth]{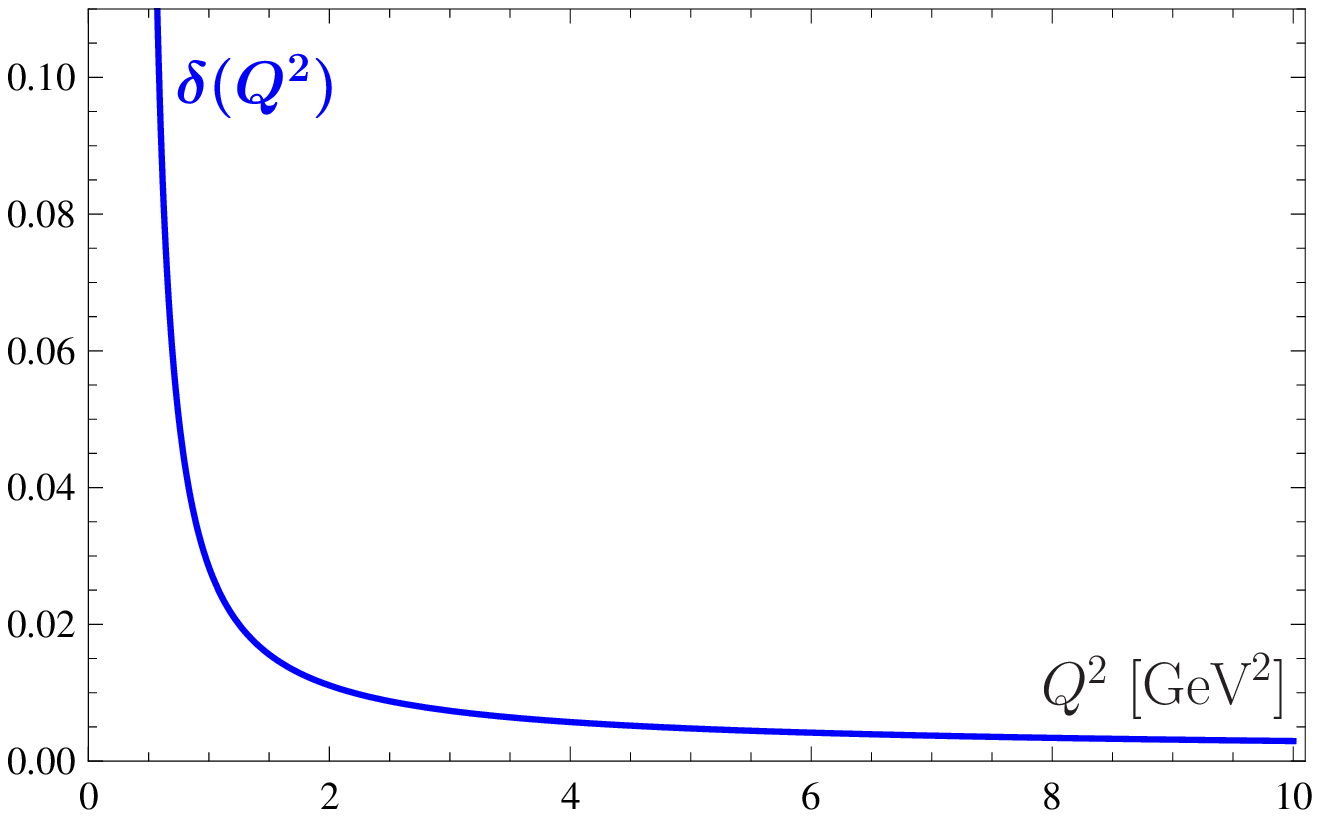}}
  \caption{Left panel: Comparison of the standard three-loop coupling
   $\alpha_s^{(3)}(Q^2)$ (solid blue line) with the three-loop Pade one
   $\alpha_s^{(3\text{P})}(Q^2)$ (dashed red line). Right panel:
   Relative accuracy $\delta(Q^2)=
   (\alpha_s^{(3\text{P})}(Q^2)-\alpha_s^{(3)}(Q^2))/\alpha_s^{(3)}(Q^2)$
   of the three-loop Pade coupling as compared with the standard three-loop one.
   \label{fig:alpha_s.3L-Pade}}
\end{figure}
As has been shown in~\cite{Mag05} this expansion
has a finite radius of convergence,
which appears to be sufficiently large for all values of $n_{f}$
of practical interest.
Note here
that this method
of expressing the higher-$\ell$-loop coupling in powers of the two-loop one
is equivalent to the 't\,Hooft scheme,
where one put by hands all coefficients in $\beta$-function,
except $b_0$ and $b_1$,
equal to zero
and effectively takes into account
all higher coefficients $b_i$
by redefining perturbative coefficients $d_i$
(see for more detail in~\cite{GaKa11}).

Another possibility for obtaining the ``exact'' three-loop solution
is provided by the so-called
Pade approximation scheme.
It is based on the Pade-type modification
of the three-loop beta function:
\begin{subequations}
 \label{eq:3L.Pade}
 \begin{eqnarray}
  \beta_{(3\text{P})}\left(\alpha_\text{s}\right)
   &=& -\frac{\alpha_\text{s}^2}{4\pi}\,
       \left[b_0
          +  \frac{b_1\,\alpha_\text{s}/(4\pi)}{1-b_2\,\alpha_\text{s}/(4\pi\,b_1)}\,
       \right]\,,
 \label{eq:betaf.3L.Pade}\\
  \frac{d a_{(3\text{P})}[L]}{d L}
   &=& -a_{(3\text{P})}^2[L]\,
       \left[1
          +  \frac{c_1\,a_{(3\text{P})}[L]}{1-c_2\,a_{(3\text{P})}[L]/c_1}\,
       \right]\,.
 \label{eq:a.3L.Pade}
 \end{eqnarray}
\end{subequations}
The last equation
can be solved exactly with the help of the same Lambert function
(here the explicit dependence on $n_f$ is not shown for shortness):
\begin{eqnarray}
 \label{eq:a.3L(P).Exact}
 a_{(3\text{P})}[L] =
   \frac{-c_1^{-1}}{1-c_2/c_1^2+W_{-1}\big(z_{W}^{(3\text{P})}[L]\big)}
 \ \ \text{with}\ \
 z_{W}^{(3\text{P})}[L]
  = -c_1^{-1}\,e^{-1+c_2/c_1^2-L/c_1}\,.
 \end{eqnarray}
The relative accuracy of this solution as compared with
numerical solution of the standard three-loop equation
(\ref{eq:beta.norm}) with $l=3$
is better than 1\% for $Q^2\geq 2$~GeV$^2$
(with $\Lambda_{3}^{(3)}=356$~MeV)
and better than $0.5\%$ for $Q^2\geq 5$~GeV$^2$,
cf. Fig.\,\ref{fig:alpha_s.3L-Pade}.

In the four-loop approximation
we use the same Eq.\,(\ref{eq:a.lL_in_2L})
with corresponding coefficients
\begin{subequations}
\begin{eqnarray}
 \label{eq:c(4)n}
  C_{n}^{(4)} = C_{n}^{(3)} +\Delta_n^{(4)}
\end{eqnarray}
and
\begin{eqnarray}
 \label{eq:del(4)n}
  \Delta_{1}^{(4)} =
  \Delta_{2}^{(4)} =
  \Delta_{3}^{(4)} = 0\,,\quad
  \Delta_{4}^{(4)} = c_3\,,\quad
  \Delta_{5}^{(4)} = \frac{-c_1\,c_3}{ 6}\,,\quad
  \Delta_{6}^{(4)} = \frac{c_1^2\,c_3}{12}+2\,c_2\,c_3\,,~~~
  \nonumber\\
  \Delta_{7}^{(4)} = \frac{-c_1^3\,c_3}{20}
                   - \frac{4\,c_1\,c_2\,c_3}{5}
                   + \frac{11\,c_3^2}{20}\,,\quad
  \Delta_{8}^{(4)} = \frac{c_1^4\,c_3}{30}
                   + \frac{9\,c_1^2\,c_2\,c_3}{20}
                   + \frac{19\,c_2^2\,c_3}{3}
                   - \frac{49\,c_1\,c_3^2}{120}\,,~~~
  \nonumber\\
  \Delta_{9}^{(4)} = \frac{c_1^5\,c_3}{42}
                   - \frac{41\,c_1^3\,c_2\,c_3}{140}
                   - \frac{946\,c_1\,c_2^2\,c_3}{315}
                   + \frac{134\,c_2\,c_3^2}{35}
                   + \frac{149\,c_1^2\,c_3^2}{504}\,.
  ~~~~~~~~~~~~~
\end{eqnarray}
\end{subequations}
\begin{figure}[t!]
 \centerline{\includegraphics[width=0.45\textwidth]{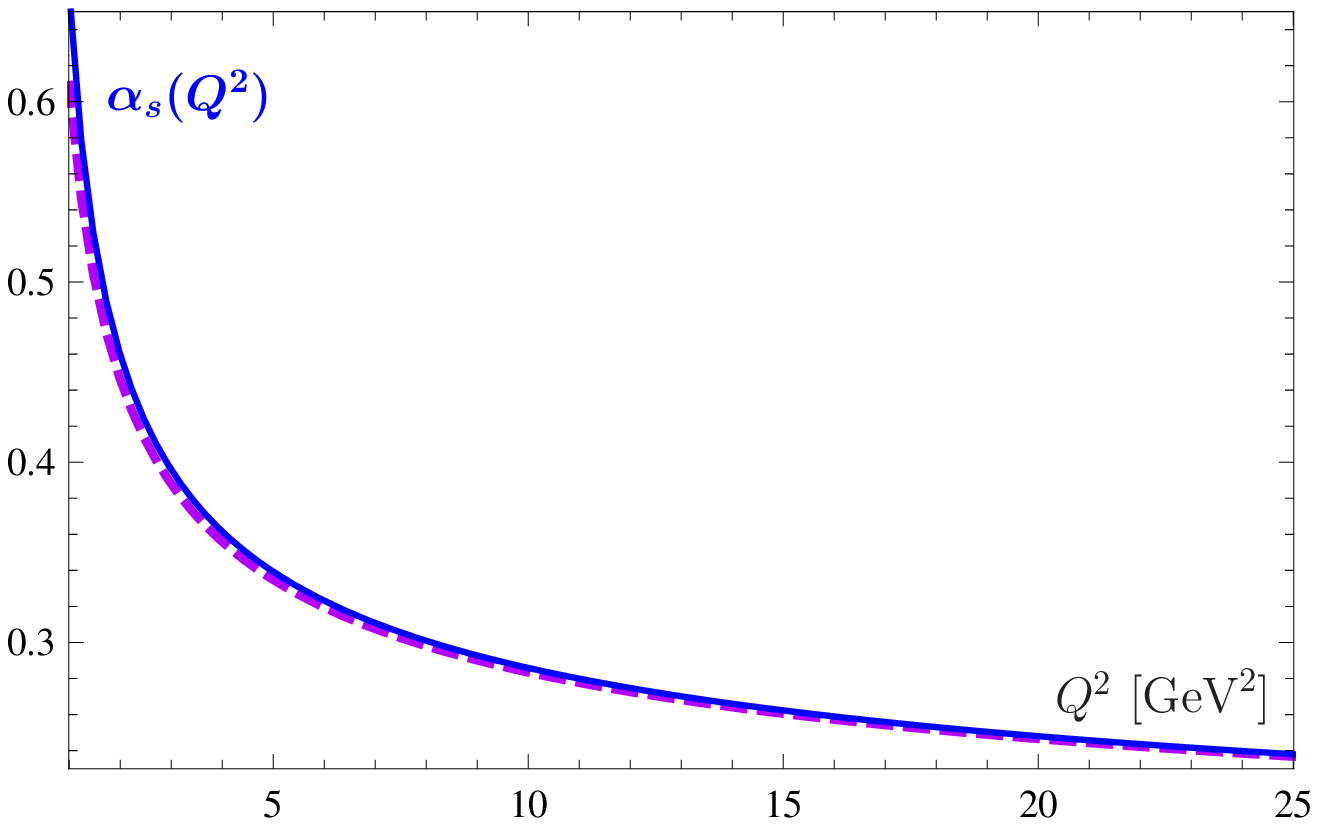}~~~
             \includegraphics[width=0.45\textwidth]{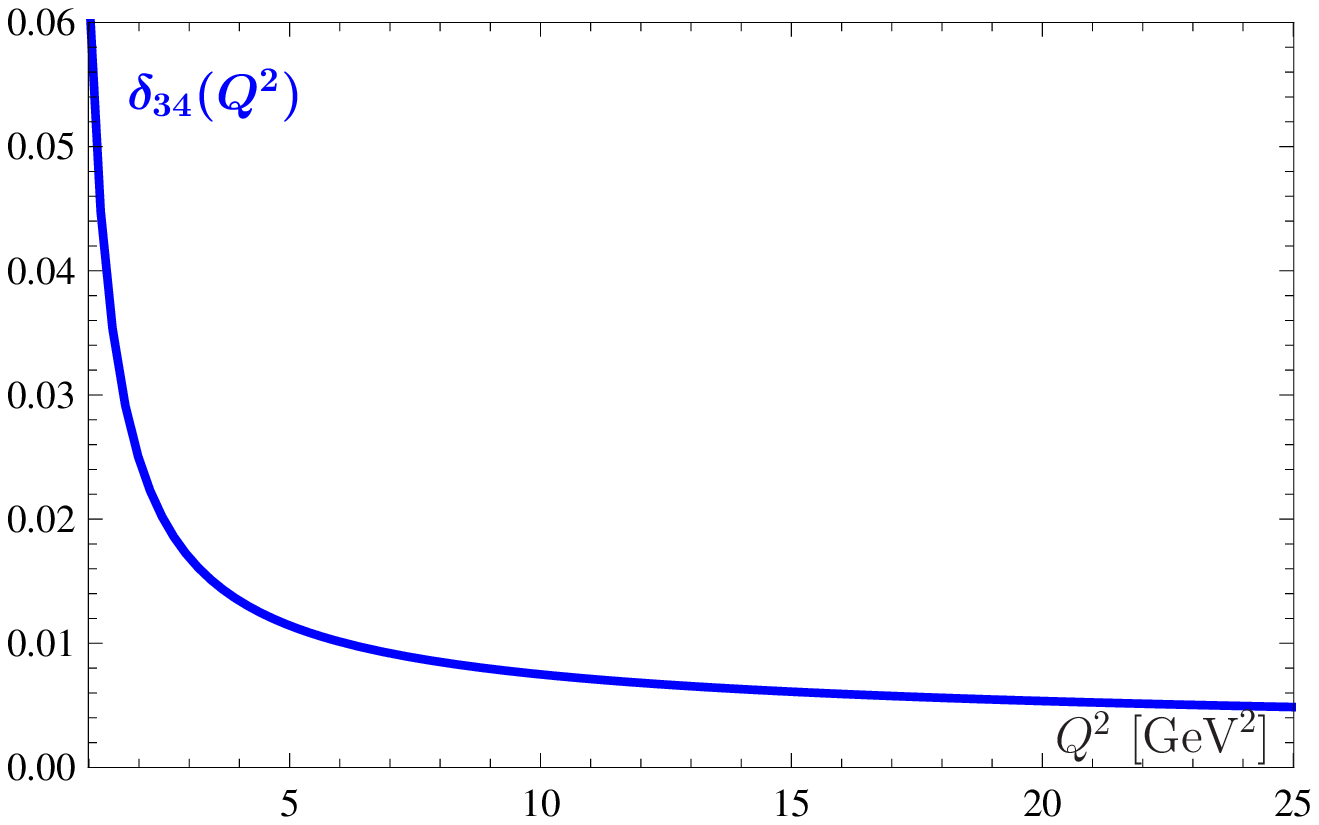}}
  \caption{Left panel: Comparison of the four-loop coupling
   $\alpha_s^{(4)}(Q^2)$ (solid blue line) with the three-loop one
   $\alpha_s^{(3)}(Q^2)$ (dashed violet line). Right panel:
   Relative accuracy $\delta_{34}(Q^2)=
   (\alpha_s^{(4)}(Q^2)-\alpha_s^{(3)}(Q^2))/\alpha_s^{(4)}(Q^2)$
   of the three-loop coupling as compared with the four-loop one.
   \label{fig:alpha_s.43L}}
\end{figure}
In the left panel of Fig.\,\ref{fig:alpha_s.43L} we show
both couplings, the four-loop
$\alpha_s^{(4)}(Q^2)$ (solid blue line),
and the three-loop
$\alpha_s^{(3)}(Q^2)$ (dashed violet line)
with fixed number of active flavors $n_f=4$.
We normalize both couplings to the same value  $\alpha_s(m_Z^2)=0.119$
at the $Z$-boson mass scale.
Numerically, as can be seen in the right panel of Fig.\,\ref{fig:alpha_s.43L},
the relative deviation
$\delta_{34}(Q^2)=(\alpha_s^{(4)}(Q^2)-\alpha_s^{(3)}(Q^2))/\alpha_s^{(4)}(Q^2)$
varies from 6\% at $Q^2=1$~GeV$^2$ to 0.5\% at $Q^2=25$~GeV$^2$.
We also compared the four-loop coupling,
calculated in accord with Eq.~(\ref{eq:a.lL_in_2L}),
with coupling,
calculated using package \texttt{RunDec}~\cite{CKS00}
with the same normalization $\alpha_s(m_Z^2)=0.119$,
--- the relative deviation appears to vary
from 0.2\% at $Q^2=1$~GeV$^2$ to 0.04\% at $Q^2=25$~GeV$^2$.

\subsection{Global scheme}
 \label{sec:pQCD.glob}
Here we consider the scheme of the so-called ``global pQCD''
in which the heavy-quark thresholds are taken into account.
We follow here to Shirkov--Solovtsov approach~\cite{SS,Mag99,Mag00}
with the following values of pole masses of $c$, $b$, and $t$ quarks:
$m_c=1.65$~GeV, $m_b=4.75$~GeV and $m_t=172.5$~GeV.
In the $\overline{\text{MS}\vphantom{^1}}$ scheme
of the standard pQCD
one needs to match the running coupling values in Euclidean domain
at $Q^2$ corresponding to these masses:
$M_4=m_c$, $M_5=m_b$ and $M_6=m_t$.
In order to implement these matching conditions
we need to use
the original QCD coupling
\begin{eqnarray}
 \label{eq:alpha.s.ell}
  \alpha_\text{s}^{(\ell)}(Q^2;n_f)
   &=& \frac{4\,\pi}{b_0(n_f)}\,
        a^{(\ell)}(Q^2;n_f)\,,
\end{eqnarray}
where the indicator $^{(\ell)}$ signals
about the loop order of the approximation we use.\footnote{
Note here
that the dependence $a^{(\ell)}(Q^2;n_f)$ on $n_f$
is the consequence of Eq.\,(\ref{eq:beta.norm}),
where for $l>1$ one has $n_f$-dependent coefficients $c_k(n_f)$.}

In what follows we use all logarithms $L$
with respect to three-flavor scale $\Lambda_3^2$:
\begin{eqnarray}
 \label{eq:L.Q2}
  L(Q^2)=\ln\left(Q^2/\Lambda_3^2\right)\,.
\end{eqnarray}
Recalculation to all other scales is realized
with the help of finite additions:
\begin{eqnarray}
 \label{eq:Log_Nf}
  \ln\left(Q^2/\Lambda_k^2\right)
    = L(Q^2)
    + \lambda_k
   \quad \text{with}
   \quad \lambda_k\equiv\ln\left(\Lambda_3^2/\Lambda_k^2\right)\,,
\end{eqnarray}
and $\Lambda_k$ --- the corresponding
to the specified value $n_f=k$ scale of QCD.
We also define the corresponding logarithmic values
at the thresholds $M_k$
($k = 4\div6$):
\begin{eqnarray}
 \label{eq:Log_Thresh_43}
  L_{k}(\Lambda_3)
   \equiv
    \ln\left(M_{k}^2/\Lambda_3^2\right)\,.
\end{eqnarray}
All QCD scales $\Lambda_f$, $f=4, 5, 6$,
we treat as functions of the single parameter,
namely, the three-flavor scale $\Lambda_3$:
\begin{eqnarray}
 \label{eq:Lambda_Nf}
 \Lambda_f \ \to\ \Lambda_f(\Lambda_3)
 \quad \text{with}\quad \Lambda_3 >
                        \Lambda_4(\Lambda_3) >
                        \Lambda_5(\Lambda_3) >
                        \Lambda_6(\Lambda_3)\,,
\end{eqnarray}
which should be defined from matching conditions
for the running coupling
at the heavy-quark thresholds.

For an illustration
we consider here the two-loop approximation
with the running coupling
$\alpha_\text{s}^{(2)}[L;n_f]$
\begin{eqnarray}
 \label{eq:Alpha.ExSol.2L}
  \alpha_\text{s}^{(2)}[L;n_f]
  = \frac{-4\,\pi}{b_0(n_f)c_1(n_f)
                    \left[1+W_{-1}(z_W[L;n_f])\right]
                   }
\end{eqnarray}
with $z_W[L;n_f]=\left(1/c_1(n_f)\right)\exp\left[-1+i\pi-L/c_1(n_f)\right]$.
Then matching conditions are
\begin{subequations}
\begin{eqnarray}
 \label{eq:matching-SS_43}
  \alpha_\text{s}^{(2)}\left[L_{4}(\Lambda_3);3\right]
  &=&
  \alpha_\text{s}^{(2)}\left[L_{4}(\Lambda_3)+\lambda_4;4\right]\,;\\
 \label{eq:matching-SS_54}
  \alpha_\text{s}^{(2)}\left[L_{5}(\Lambda_3)+\lambda_4;4\right]
  &=&
  \alpha_\text{s}^{(2)}\left[L_{5}(\Lambda_3)+\lambda_5;5\right]\,;\\
 \label{eq:matching-SS_65}
  \alpha_\text{s}^{(2)}\left[L_{6}(\Lambda_3)+\lambda_5;5\right]
  &=&
  \alpha_\text{s}^{(2)}\left[L_{6}(\Lambda_3)+\lambda_6;6\right]\,.
\end{eqnarray}
\end{subequations}
These relations define constants $\lambda_k$ with $k = 4\div6$
as functions of variable $\Lambda_3$,
namely
\begin{eqnarray}
 \label{eq:lam.f(Lam3)}
  \lambda_k \to \lambda_k^{(2)}(\Lambda_3)\,,
\end{eqnarray}
and,
as a consequence,
the continuous global effective QCD coupling
\begin{eqnarray}
 \label{eq:Alpha.Glob.2L}
  \alpha_\text{s}^{\text{glob};(2)}(Q^2,\Lambda_3)
   \!&\!=\!&\! \alpha_\text{s}^{(2)}\left[L(Q^2);3\right]
           \theta\left(Q^2\!<\!M_4^2\right)
\nonumber\\
   \!&\!+\!&\! \alpha_\text{s}^{(2)}\left[L(Q^2)\!+\!\lambda_4^{(2)}(\Lambda_3);4\right]
         \theta\left(M_4^2\!\leq\!Q^2<\!M_5^2\right)
\nonumber\\
   \!&\!+\!&\! \alpha_\text{s}^{(2)}\left[L(Q^2)\!+\!\lambda_5^{(2)}(\Lambda_3);5\right]
         \theta\left(M_5^2\!\leq\!Q^2<\!M_6^2\right)
\nonumber\\
   \!&\!+\!&\! \alpha_\text{s}^{(2)}\left[L(Q^2)\!+\!\lambda_6^{(2)}(\Lambda_3);6\right]
         \theta\left(M_6^2\!\leq\!Q^2\right).~~~
\end{eqnarray}
Here is the list of partial values
of $\Lambda_f^{(2)}(\Lambda_3)$, $\lambda_f^{(2)}(\Lambda_3)$ and $L_{f}(\Lambda_3)$
with $f=4, 5, 6$ for $\Lambda_3=400$ MeV:
\begin{subequations}
\begin{eqnarray}
 \label{eq:Lambda_f.400MeV}
  \Lambda_4^{(2)} &=& 333~\text{MeV}\,,\quad
  \Lambda_5^{(2)}\ =\ 233~\text{MeV}\,,\quad
  \Lambda_6^{(2)}\ =\ 98~\text{MeV}\,;\\
 \label{eq:lambda_f.400MeV}
  \lambda_4^{(2)} &=& 0.367\,,~~~~~~~~
  \lambda_5^{(2)}\ =\ 1.08\,,~~~~~~~~~~
  \lambda_6^{(2)}\ =\ 2.82\,;\\
 \label{eq:L_f3.400MeV}
  L_{4} &=& 2.197\,,~~~~~~~~~
  L_{5}\ =\ 4.750\,,~~~~~~~~~~
  L_{6}\ =\ 12.162\,.
\end{eqnarray}
\end{subequations}

In our m-file we use the following realizations.
The QCD scales are encoded as
$\Lambda1[\Lambda, n_f]$, $\Lambda2[\Lambda, n_f]$,
and $\Lambda3[\Lambda, n_f]$
(in \texttt{Mathematica} capital Greek symbol $\Lambda$
 can be written as \verb'\[CapitalLambda]'):
\begin{subequations}
\begin{eqnarray}
 \label{eq:Lambda.f}
  \verb'\[CapitalLambda]'{\ell}[\Lambda, k]
  = \Lambda{\ell}[\Lambda, n_f=k]
  = \Lambda_k^{(\ell)}(\Lambda)\,,\,
  (\ell = 1\div4, 3\text{P}\,;\, k=4\div6)\,,~
\end{eqnarray}
the threshold logarithms --- as
$\lambda{\ell}{4}[\Lambda]$, $\lambda{\ell}{5}[\Lambda]$,
and $\lambda{\ell}{6}[\Lambda]$
(in \texttt{Mathematica} Greek symbol $\lambda$
 can be written as \verb'\[Lambda]'):
\begin{eqnarray}
 \label{eq:Lambda.lNf}
  \verb'\[Lambda]'{\ell}{k}[\Lambda]
   = \lambda{\ell}{k}[\Lambda]
   = \ln\left(\Lambda^2/\Lambda{\ell}[\Lambda,k]^2\right)\,,\quad
  (\ell = 1\div4, 3\text{P}\,;\ k = 4\div6)\,,
\end{eqnarray}
the running QCD couplings with fixed $n_f$ --- as
$\alpha\text{Bar}1[Q^2,n_f,\Lambda]$,
$\alpha\text{Bar}2[Q^2,n_f,\Lambda]$, and
$\alpha\text{Bar}3[Q^2,n_f,\Lambda]$
(in \texttt{Mathematica} Greek symbol $\alpha$
 can be written as \verb'\[Alpha]'):
\begin{eqnarray}
 \label{eq:AlphaBarl}
  \verb'\[Alpha]'\text{Bar}{\ell}[Q^2,n_f,\Lambda]
  = \alpha\text{Bar}{\ell}[Q^2,n_f,\Lambda]
  = \alpha_\text{s}^{(\ell)}[\ln(Q^2/\Lambda^2);n_f],\,
  (\ell = 1\div4, 3\text{P})\,,~
\end{eqnarray}
and the global running QCD couplings --- as
$\alpha\text{Glob}1[Q^2,\Lambda]$,
$\alpha\text{Glob}2[Q^2,\Lambda]$, and
$\alpha\text{Glob}3[Q^2,\Lambda]$:
\begin{eqnarray}
 \label{eq:AlphaGlobl}
  \verb'\[Alpha]'\text{Glob}{\ell}[Q^2,\Lambda]
   = \alpha\text{Glob}{\ell}[Q^2,\Lambda]
   = \alpha_\text{s}^{\text{glob};(\ell)}(Q^2,\Lambda)\,,\,
  (\ell = 1\div4, 3\text{P})\,,~
\end{eqnarray}
\end{subequations}

To be more specific,
we consider here  an example.
We assume that the two-loop $\alpha_\text{s}$ is given at the $Z$-boson scale
as $\alpha_\text{s}^{(2)}[\ln(m_Z^2/\Lambda^2);5]=0.119$.
We want to evaluate the corresponding values of
the QCD scales $\Lambda_3$, $\Lambda_4$, and $\Lambda_5$
and the coupling
$\alpha_\text{s}^{\text{glob};(\ell)}(Q^2,\Lambda)$
at the scale $Q=M_5$.
We show a possible \texttt{Mathematica} realization of this task.
\begin{verbatim}
In[1]:= <<FAPT.m;
\end{verbatim}
Comment: \texttt{NumDefFAPT} is a set of \texttt{Mathematica} rules
in our package
which assigns typical values to the physical parameters
used in our procedures.
\begin{verbatim}
In[2]:= {MZ = MZboson/.NumDefFAPT, Mb=MQ5/.NumDefFAPT}
Out[2]= {91.19, 4.75}
\end{verbatim}
Comment: evaluation of L23$=\Lambda^{(2)}_3$ from $\alpha_\text{s}^{(2)}[\ln(m_Z^2/\Lambda^2);5]$
based on the explicit solution, Eq.~(\ref{eq:Alpha.ExSol.2L}), Eq.\,(\ref{eq:Alpha.Glob.2L}).
\begin{verbatim}
In[3]:= L23=lx/.FindRoot[\[Alpha]Glob2[MZ^2,lx]==0.119, {lx,0.1,0.3}]
Out[3]= 0.387282
\end{verbatim}
Comment: evaluation of L24$=\Lambda_4^{(2)}$ and L25$=\Lambda_5^{(2)}$
from L23$=\Lambda_3^{(2)}$
based on Eq.\,(\ref{eq:Lambda.f}).
\begin{verbatim}
In[4]:= {L24=\[CapitalLambda]2[L23,4], L25=\[CapitalLambda]2[L23,5]}
Out[4]= {0.321298, 0.224033}
\end{verbatim}
Comment: evaluation of $\alpha_\text{s}^{\text{glob};(2)}(M_b^2)$
from L23$=\Lambda_3^{(2)}$.
\begin{verbatim}
In[5]:= \[Alpha]Glob2[Mb^2,L23]
Out[5]= 0.218894
\end{verbatim}

\section{Basics of FAPT}
 \label{sec:FAPT}
In the end of the previous section
we used for the running QCD couplings with fixed $n_f$
the \texttt{Bar} notations ---
$\alpha\text{Bar}1[Q^2,n_f,\Lambda]$,
$\alpha\text{Bar}2[Q^2,n_f,\Lambda]$, and
$\alpha\text{Bar}3[Q^2,n_f,\Lambda]$.
We did it on purpose
to have a direct connection to our previous papers on the
subject \cite{BMS05,BKS05,BMS06,AB08},
where
we used the normalized coupling
$a(\mu^2)=\beta_f\,\alpha_\text{s}(\mu^2)$,
cf. Eq.\,(\ref{eq:ck.def}).
To be in line with these definitions,
we also introduce analogous expressions
for the fixed-$N_f$ quantities
with standard normalization,
i.e.,
\begin{eqnarray}
 \bar{\mathcal A}_{\nu}(Q^2)
=
  \frac{{\mathcal A}_{\nu}(Q^2)}{\beta_{f}^{\nu}}\, , \quad
  \bar{\mathfrak A}_{\nu}(s)
=
  \frac{{\mathfrak A}_{\nu}(s)}{\beta_{f}^{\nu}}\, ,
\label{eq:Bar.Couplings}
\end{eqnarray}
which correspond to the analytic couplings
${\mathcal A}_{\nu}$ and ${\mathfrak A}_{\nu}$
in the Shirkov--Solovtsov terminology \cite{SS}.

The basic objects in the (F)APT approach are spectral densities
{${\bar\rho_{\nu}^{(\ell)}}(\sigma;n_f)$}
which enter the K\"allen--Lehmann spectral representation
for the analytic couplings:
\begin{subequations}
 \label{eq:A.U}
 \begin{eqnarray}
  \label{eq:A_1}
   \bar{\mathcal A}_\nu^{(\ell)}[L;n_f]
   \!&\!=\!&\! \int_0^{\infty}\!\frac{\bar\rho_\nu^{(\ell)}(\sigma;n_f)}{\sigma+Q^2}\,
                d\sigma
        =      \int_{-\infty}^{\infty}\!\frac{\bar\rho_\nu^{(\ell)}[L_\sigma;n_f]}{1+\exp(L-L_\sigma)}\,
                dL_\sigma\,,\\
  \label{eq:U_1}
   \bar{\mathfrak A}_\nu^{(\ell)}[L_s;n_f]
   \!&\!=\!&\! \int_s^{\infty}\!\frac{\bar\rho_\nu^{(\ell)}(\sigma;n_f)}{\sigma}\,
                d\sigma
            = \int_{L_s}^{\infty}\!\bar\rho_\nu^{(\ell)}[L_\sigma;n_f]\,
               dL_\sigma\,,
 \end{eqnarray}
\end{subequations}
It is convenient to use the following representation
for spectral functions
\begin{eqnarray}
 \bar\rho_{\nu}^{(\ell)}[L;n_f]
  = \frac{1}{\pi}\,
     \textbf{Im}{}
      \left(\alpha_\text{s}^{(\ell)}\left[L-i\pi;n_f\right]
      \right)^{\nu}
  = \frac{\sin[\nu\,\varphi_{(\ell)}[L;n_f]]}
         {\pi\,(\beta_f\,R_{(\ell)}[L;n_f])^{\nu}}\,,
 \label{eq:SpDen.lL.nu}
\end{eqnarray}
which is based on the module-phase representation of a complex number
\begin{eqnarray}
 \label{eq:a(l).R.phi}
  \alpha_\text{s}^{(\ell)}\left[L-i\pi;n_f\right]
   = \frac{a_{(\ell)}\left[L-i\pi;n_f\right]}{\beta_f(n_f)}
   = \frac{\displaystyle e^{i\varphi_{(\ell)}[L;n_f]}}{\beta_f(n_f)\,R_{(\ell)}[L;n_f]}\,.
\end{eqnarray}
In the one-loop approximation the corresponding functions
have the most simple form
\begin{eqnarray}
 \varphi_{(1)}[L]
  = \arccos\left(\frac{L}{\sqrt{L^2+\pi^2}}\right)\,,~~
 R_{(1)}[L]
  = \sqrt{L^2+\pi^2}
 \label{eq:SpDen.1L.n}
\end{eqnarray}
and do not depend on $n_f$,
whereas at the two-loop order
they have a more complicated form
\begin{subequations}
\label{eq:SpDen.nu.(2).R.phi}
\begin{eqnarray}
 \label{eq:Lamb.R_(2)}
  R_{(2)}[L;n_f]
   &=& c_1(n_f)\,\Big|1+W_{1}\left(z_W[L-i\pi;n_f]\right)\Big|\,,\\
 \label{eq:Lamb.phi_(2)}
  \varphi_{(2)}[L;n_f]
   &=& \arccos
       \left[
        \textbf{Re}\left(
                   \frac{-R_{(2)}[L;n_f]}
                        {1+W_{1}\left(z_W[L-i\pi;n_f]\right)}
                   \right)
      \right]
\end{eqnarray}
\end{subequations}
with $W_{1}[z]$ being the appropriate branch of Lambert function.
In the three-loop approximation we use
either Eq.\,(\ref{eq:a.lL_in_2L})
and then obtain
\begin{subequations}
\begin{eqnarray}
 \label{eq:Lamb.R_(3)}
  R_{(3)}[L]
   &=& \left|\frac{e^{i\,\varphi_{(2)}[L]}}{R_{(2)}[L]}
             + \sum_{k\geq3}
                C_{k}^{(3)}\,\frac{e^{i\,k\,\varphi_{(2)}[L]}}{R_{(2)}^k[L]}
       \right|^{-1}\,;~~~\\
 \label{eq:Lamb.phi_(3)}
  \varphi_{(3)}[L]
   &=& \arccos
      \left[\frac{R_{(3)}[L]\cos\left(\varphi_{(2)}[L]\right)}
                 {R_{(2)}[L]}
         + \sum_{k\geq3}
            C_{k}^{(3)}\,
             \frac{R_{(3)}[L]\cos\left(k\,\varphi_{(2)}[L]\right)}
                  {R_{(2)}^k[L]}
      \right]
\end{eqnarray}
\end{subequations}
or Eq.\,(\ref{eq:a.3L(P).Exact})
--- and then obtain
\begin{subequations}
\begin{eqnarray}
 \label{eq:Lamb.R_(3P)}
  R_{(3\text{P})}[L]
   &=& c_1\,\bigg|1-\frac{c_2}{c_1^2}+W_{1}\left(z_{W}^{(3\text{P})}[L-i\pi]\right)\bigg|\,;\\
 \label{eq:Lamb.phi_(3P)}
  \varphi_{(3\text{P})}[L]
   &=& \arccos
        \left[
         \textbf{Re}
          \left(\frac{-R_{(3\text{P})}[L]}
                     {1-(c_2/c_1^2)+W_{1}\left(z_{W}^{(3\text{P})}[L-i\pi]\right)}
          \right)
        \right]\,.
\end{eqnarray}
\end{subequations}
In the four-loop approximation we use
Eq.\,(\ref{eq:a.lL_in_2L})
and then obtain
\begin{subequations}
\begin{eqnarray}
 \label{eq:Lamb.R_(4)}
  R_{(4)}[L]
   &=& \left|\frac{e^{i\,\varphi_{(2)}[L]}}{R_{(2)}[L]}
             + \sum_{k\geq3}
                C_{k}^{(4)}\,\frac{e^{i\,k\,\varphi_{(2)}[L]}}{R_{(2)}^k[L]}
       \right|^{-1}\,;~~~\\
 \label{eq:Lamb.phi_(4)}
  \varphi_{(4)}[L]
   &=& \arccos
      \left[\frac{R_{(4)}[L]\cos\left(\varphi_{(2)}[L]\right)}
                 {R_{(2)}[L]}
         + \sum_{k\geq3}
            C_{k}^{(4)}\,
             \frac{R_{(4)}[L]\cos\left(k\,\varphi_{(2)}[L]\right)}
                  {R_{(2)}^k[L]}
      \right]
\end{eqnarray}
\end{subequations}
Here we do not show explicitly the $n_f$ dependence
of the corresponding quantities ---
it goes inside through
$R_{(2)}[L]=R_{(2)}[L;n_f]$,
$\varphi_{(2)}[L]=\varphi_{(2)}[L;n_f]$,
$C_{k}^{(3)}=C_{k}^{(3)}[n_f]$,
$C_{k}^{(4)}=C_{k}^{(4)}[n_f]$,
$c_{k}=c_{k}(n_f)$
with $k=1\div3$,
and
$z_{W}^{(3\text{P})}[L]=z_{W}^{(3\text{P})}[L;n_f]$.
\begin{figure}[h!]
 \centerline{\includegraphics[width=0.45\textwidth]{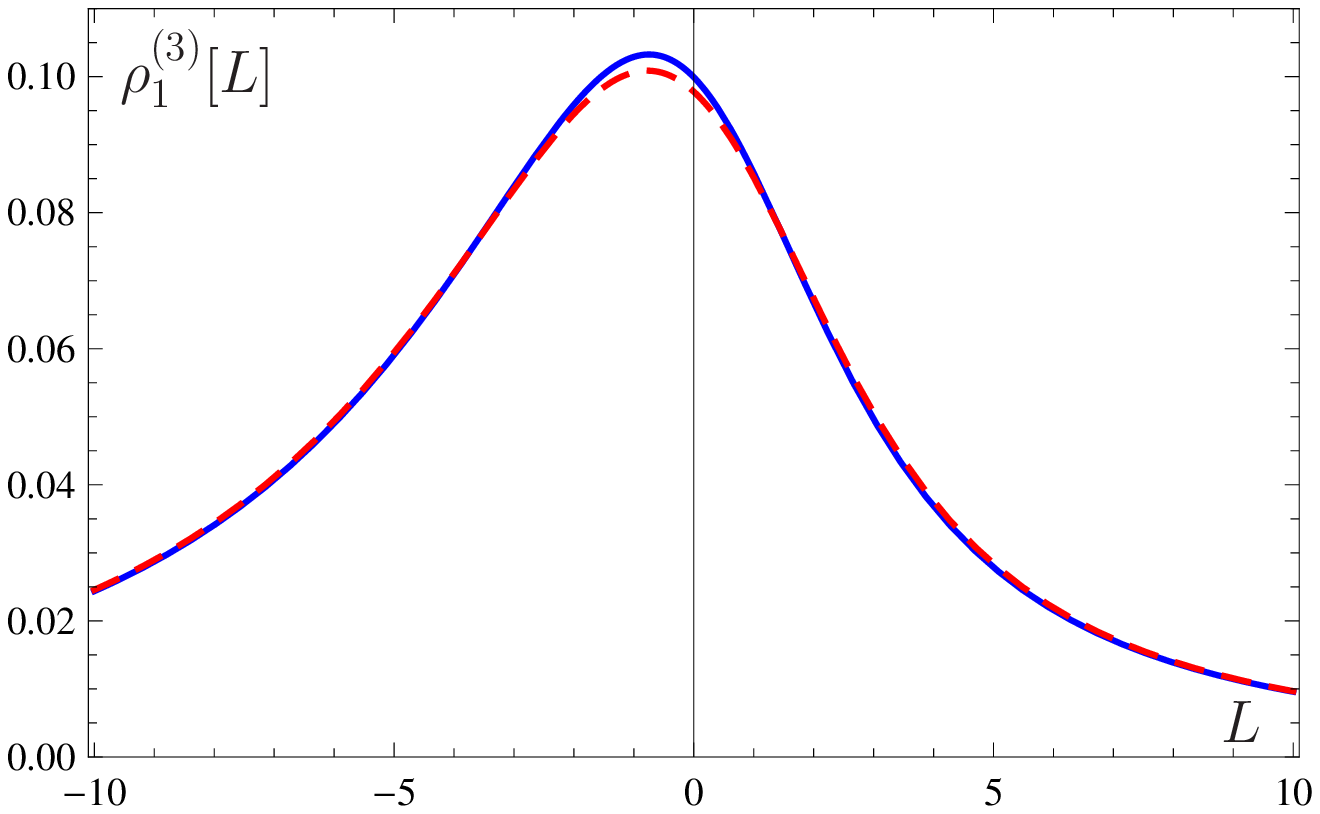}~~~
             \includegraphics[width=0.45\textwidth]{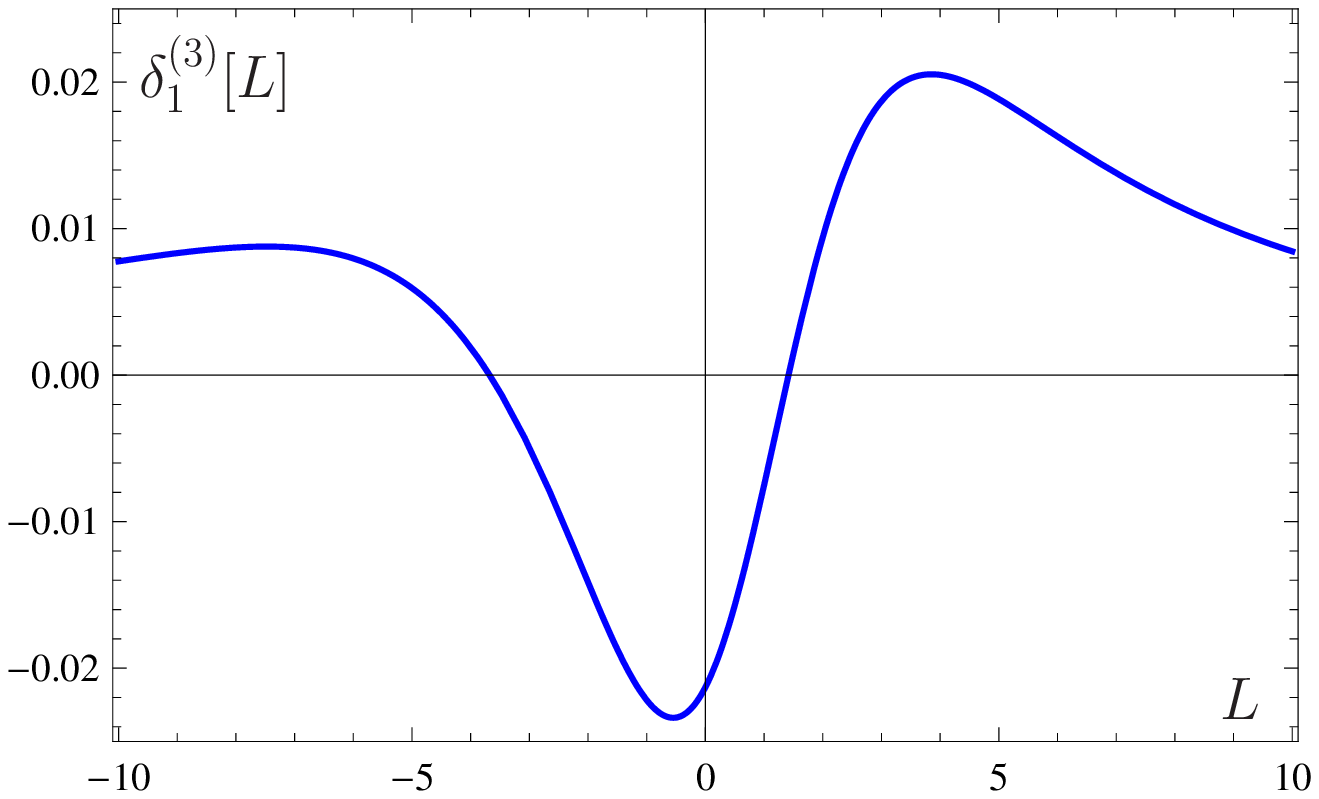}}
  \caption{Left panel: Comparison of the standard three-loop spectral density
   $\rho_1^{(3)}[L]$ (solid blue line) with the three-loop Pade one
   $\rho_1^{(3\text{P})}[L]$ (dashed red line). Right panel:
   Relative accuracy $\delta_1^{(3)}[L]=
   (\rho_1^{(3\text{P})}[L]-\rho_1^{(3)}[L])/\rho_1^{(3)}[L]$
   of the three-loop Pade spectral density as compared
   with the standard three-loop one.
   \label{fig:rho.3L-Mag-Pade}}
\end{figure}
In the left panel of Fig.\,\ref{fig:rho.3L-Mag-Pade}
we show both spectral densities
in comparison.
On the right panel of this figure
we show the relative deviation
of the Pade spectral density from the standard one:
one can see that it varies from $+1$\% at $L\approx -7$,
reduces to $-2$\% at $L\approx0$,
and then reaches the maximum of $+2$\% at $L\approx 3.5$.

  In accordance with Eq.\,(\ref{eq:Alpha.Glob.2L})
the global spectral densities are constructed through the $n_f$-fixed ones
in the following manner:
\begin{eqnarray}
 \label{eq:Rho[L].Glo.n}
  \rho_{\nu}^{(\ell);\text{glob}}[L_{\sigma},\Lambda_3]
  &=& \bar{\rho}_{\nu}^{(\ell)}\left[L_{\sigma};3\right]
       \theta\left(L_{\sigma}<L_{4}(\Lambda_3)\right)
    + \bar{\rho}_{\nu}^{(\ell)}\left[L_{\sigma}+\lambda_6^{(\ell)}(\Lambda_3);6\right]\,
       \theta\left(L_{6}(\Lambda_3)\leq L_{\sigma}\right)
  \nonumber\\
  &+& \bar{\rho}_{\nu}^{(\ell)}\left[L_{\sigma}+\lambda_4^{(\ell)}(\Lambda_3);4\right]\,
       \theta\left(L_{4}(\Lambda_3)\leq L_{\sigma}<L_{5}(\Lambda_3)\right)
  \nonumber\\
  &+& \bar{\rho}_{\nu}^{(\ell)}\left[L_{\sigma}+\lambda_5^{(\ell)}(\Lambda_3);5\right]\,
       \theta\left(L_{5}(\Lambda_3)\leq L_{\sigma}<L_{6}(\Lambda_3)\right)
\end{eqnarray}
with $L_{\sigma}\equiv\ln(\sigma/\Lambda_3^2)$
and the corresponding global analytic couplings
are
\begin{subequations}
 \label{eq:A.U.Glo}
 \begin{eqnarray}
  \label{eq:A.Glo}
   \mathcal A_\nu^{(\ell);\text{glob}}[L,\Lambda_3]
   &=& \int_{-\infty}^{\infty}\!
        \frac{\rho_\nu^{(\ell);\text{glob}}[L_\sigma,\Lambda_3]}
             {1+\exp(L-L_\sigma)}\,
         dL_\sigma\,,\\
  \label{eq:U.Glo}
   \mathfrak A_\nu^{(\ell);\text{glob}}[L,\Lambda_3]
   &=& \int_{L}^{\infty}\!
        \rho_\nu^{(\ell);\text{glob}}[L_\sigma,\Lambda_3]\,
         dL_\sigma\,,
 \end{eqnarray}
\end{subequations}

\section{FAPT Procedures}
 \label{sec:proc}

In our package \texttt{FAPT.m} we use the following realizations
for the spectral densities.
$\text{RhoBar}\ell[L, n_f, \nu]$ returns $\ell$-loop spectral
density ${\bar\rho_{\nu}^{(\ell)}}$ $(\ell=1, 2, 3, 3\text{P}, 4)$ of
fractional-power $\nu$ at $L=\ln(Q^2/\Lambda^2)$
and at fixed number of active quark flavors $n_f$:
\begin{subequations}
\begin{eqnarray}
 \label{eq:RhoBar.f}
  \verb'RhoBar'{\ell}[L, k, \nu]
  = \bar\rho_{\nu}^{(\ell)}[L;n_f=k]\,,\quad
  (\ell = 1\div4, 3\text{P}\,;\ k = 3\div6)\,,~~~
\end{eqnarray}
whereas
$\text{RhoGlob}\ell[L, \nu, \Lambda_3]$ returns the global
$\ell$-loop spectral density ${\bar\rho_{\nu}^{(\ell);\text{glob}}}[L;\Lambda_3]$
$(\ell=1, 2, 3, 3\text{P}, 4)$
of fractional-power $\nu$
at $L=\ln(Q^2/\Lambda_3^2)$,
cf.
and with $\Lambda_3$ being the QCD $n_f=3$-scale:
\begin{eqnarray}
 \label{eq:RhoBar.glob}
  \verb'RhoGlob'{\ell}[L, \nu, \Lambda_3]
  = \bar\rho_{\nu}^{(\ell);\text{glob}}[L;\Lambda_3]\,,\quad
  (\ell = 1\div4, 3\text{P})\,,~~~
\end{eqnarray}
\end{subequations}

Analogously,
$\text{AcalBar}\ell[L, n_f, \nu]$
returns $\ell$-loop $(\ell=1, 2, 3, 3\text{P}, 4)$
analytic image
of fractional-power $\nu$
coupling $\bar{\mathcal A}_{\nu}^{(\ell)}[L;n_f]$
in Euclidean domain,
\begin{subequations}
\label{eq:AUcalBar.nf}
\begin{eqnarray}
 \label{eq:AcalBar.nf}
  \verb'AcalBar'{\ell}[L,k,\nu]
  = \bar{\mathcal A}_{\nu}^{(\ell)}[L;n_f=k]\,,\quad
  (\ell = 1\div4, 3\text{P}\,;\ k = 3\div6)\,,~~~
\end{eqnarray}
and
$\text{UcalBar}\ell[L, n_f, \nu]$
returns $\ell$-loop $(\ell=1, 2, 3, 3\text{P}, 4)$
analytic image of
fractional-power $\nu$ coupling
$\bar{\mathfrak A}_{\nu}^{(\ell)}[L,n_f]$
in Minkowski domain,
\begin{eqnarray}
 \label{eq:UcalBar.nf}
  \verb'UcalBar'{\ell}[L,k,\nu]
  = \bar{\mathfrak A}_{\nu}^{(\ell)}[L;n_f=k]\,,\quad
  (\ell = 1\div4, 3\text{P}\,;\ k = 3\div6)\,,~~~
\end{eqnarray}
\end{subequations}

In global case
$\text{AcalGlob}\ell[L, \nu, \Lambda_3]$
returns $\ell$-loop analytic image of fractional-power $\nu$ coupling
${\mathcal A}_{\nu}^{(\ell);glob}[L,\Lambda_3]$ in
Euclidean domain,
\begin{subequations}
\label{eq:AUcalGlob}
\begin{eqnarray}
 \label{eq:AcalGlob}
  \verb'AcalGlob'{\ell}[L,\nu,\Lambda_3]
  = \mathcal A_{\nu}^{(\ell);\text{glob}}[L,\Lambda_3]\,,\quad
  (\ell = 1\div4, 3\text{P})\,,~~~
\end{eqnarray}
and
$\text{UcalGlob}\ell[L,\nu,\Lambda_3]$
returns $\ell$-loop analytic image of fractional-power $\nu$ coupling
$\mathfrak A_{\nu}^{(\ell);glob}[L,\Lambda_3]$
in Minkowski domain,
\begin{eqnarray}
 \label{eq:UcalGlob}
  \verb'UcalGlob'{\ell}[L,\nu,\Lambda_3]
  = \mathfrak A_{\nu}^{(\ell);glob}[L,\Lambda_3]\,,\quad
  (\ell = 1\div4, 3\text{P})\,.
\end{eqnarray}
\end{subequations}

We consider here an example of using this quantities
in case of
\texttt{Mathematica}~7.
We assume that the two-loop QCD scale $\Lambda_3$
is fixed at the value $\Lambda_3=0.387$~GeV
defined at the end of section~\ref{sec:pQCD.glob}.
\begin{verbatim}
In[1]:= <<FAPT.m;
\end{verbatim}
\begin{verbatim}
In[2]:= L23=0.387;
\end{verbatim}
We determine the value
of the two-loop QCD scale $\text{L23APT}=\Lambda_3^{(2);\text{\tiny APT}}$
in APT,
corresponding to the same value 0.119 as before,
but now for the global analytic coupling:
\begin{verbatim}
In[3]:= MZ = MZboson /. NumDefFAPT
Out[3]= 91.19

In[4]:= L23APT=lx/.FindRoot[AcalGlob2[Log[MZ^2/lx^2],1,lx]
                            == 0.119,{lx,0.35,0.45}];
Out[4]= 0.379788
\end{verbatim}
Now we evaluate the value of
$\mathcal A_{\nu}^{(2);\text{glob}}[L,\text{L23APT}]$
for $L=-5.0$, $-3.0$, $-1.0$, $1.0$, $3.0$, $5.0$
with indication of the needed time:
\begin{verbatim}
In[5]:= {L0=-5., AcalGlob2[L0,1,L23APT]}//Timing
Out[5]= {0.734, {-5., 0.929485}}
\end{verbatim}
\begin{verbatim}
In[6]:= {L0=-3., AcalGlob2[L0,1,L23APT]}//Timing
Out[6]= {0.421, {-3.,0.786904}}
\end{verbatim}
\begin{verbatim}
In[7]:= {L0=-1., AcalGlob2[L0,1,L23APT]}//Timing
Out[7]= {0.422, {-1.,0.60986}}
\end{verbatim}
\begin{verbatim}
In[8]:= {L0=1., AcalGlob2[L0,1,L23APT]}//Timing
Out[8]= {0.437, {1.,0.434041}}
\end{verbatim}
\begin{verbatim}
In[9]:= {L0=3., AcalGlob2[L0,1,L23APT]}//Timing
Out[9]= {0.469, {3.,0.301442}}
\end{verbatim}
\begin{verbatim}
In[10]:= {L0=5., AcalGlob2[L0,1,L23APT]}//Timing
Out[10]= {0.531, {5.,0.219137}}
\end{verbatim}
Now we create a two-dimensional plot of
$\mathcal A_{\nu}^{(2);\text{glob}}[L,\text{L23APT}]$
and
$\mathfrak A_{\nu}^{(2);\text{glob}}[L,\text{L23APT}]$
for $L\in[-3,11]$
with indication of the needed time:
\begin{verbatim}
In[11]:= Plot[AcalGlob2[L,1,L23APT],{L,-3,11},MaxRecursion->1]//Timing
Out[11]= {19.843, Graphics (see in the left panel of Fig.\,4)}
\end{verbatim}
\begin{verbatim}
In[12]:= Plot[UcalGlob2[L,1,L23APT],{L,-3,11},MaxRecursion->1]//Timing
Out[12]= {14.656, Graphics (see in the right panel of Fig.\,4)}
\end{verbatim}
\begin{figure}[t!]
 \centerline{\includegraphics[width=0.45\textwidth]{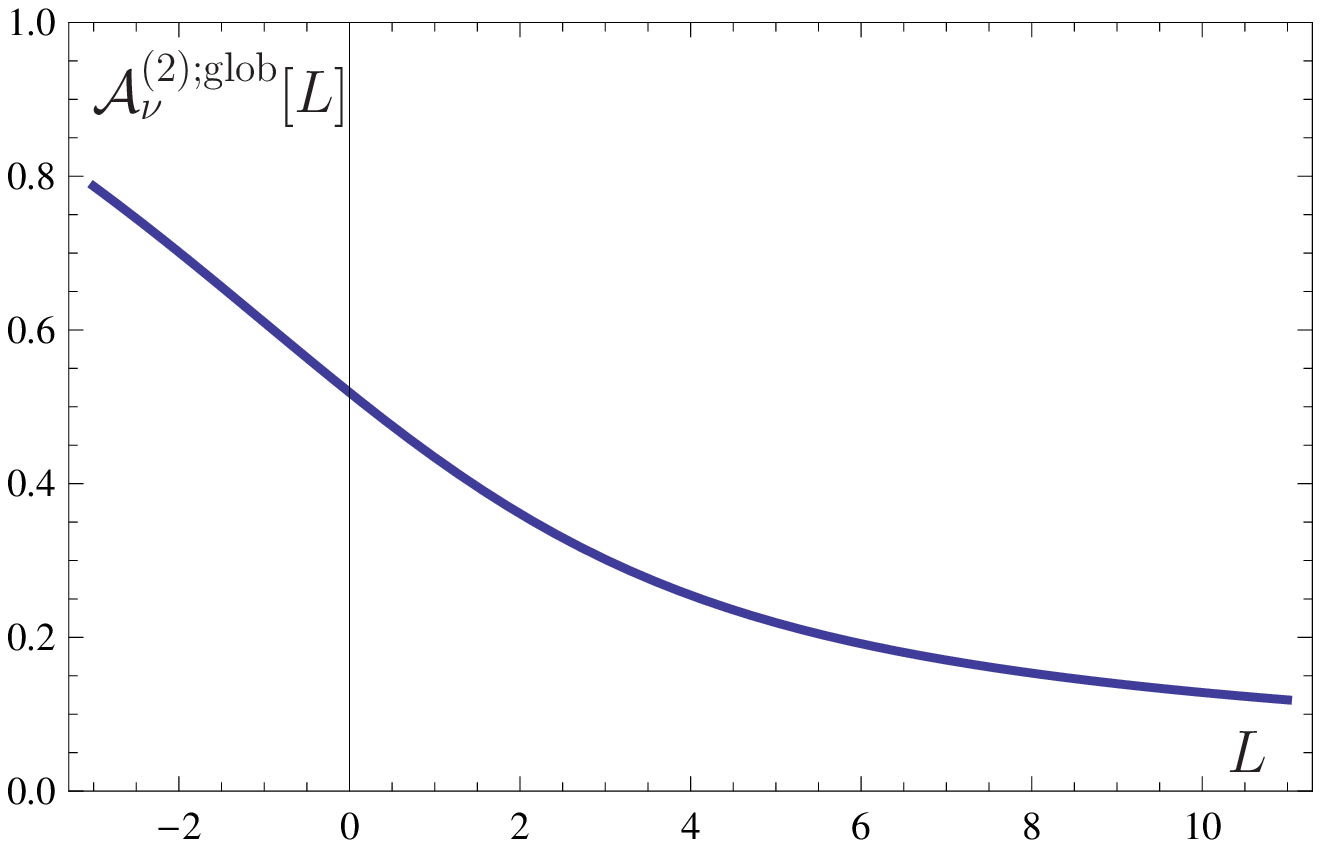}~~~
             \includegraphics[width=0.45\textwidth]{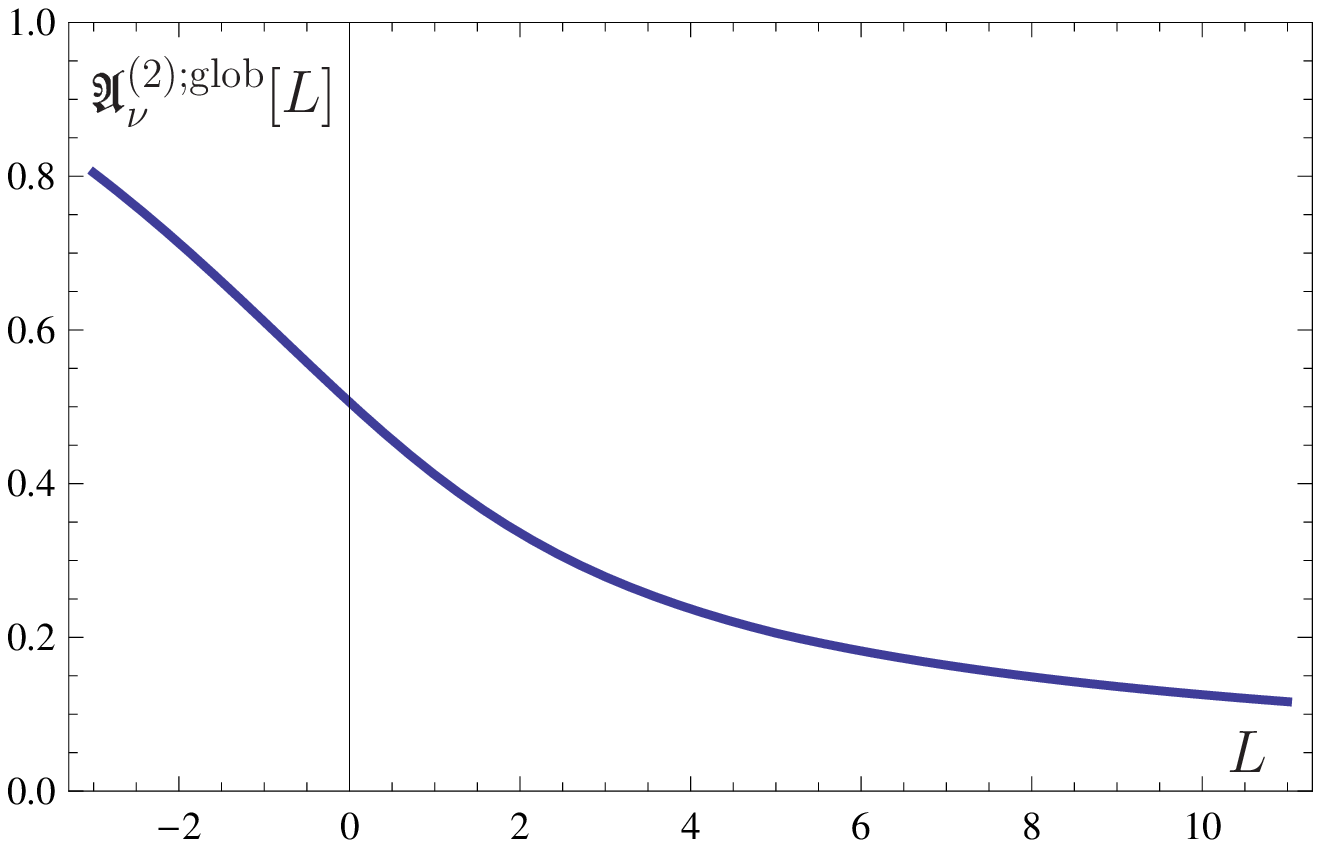}}
  \caption{Left panel: Graphics produced in Out[10] for
   $\mathcal A_{\nu}^{(2);\text{glob}}[L,\text{L23APT}]$ as a function of $L$.
   Right panel: Graphics produced in Out[11] for
   $\mathfrak A_{\nu}^{(2);\text{glob}}[L,\text{L23APT}]$ as a function of $L$.
   \label{fig:AU.2L}}
\end{figure}

\section{Interpolation}
 \label{sec:Interpol}

The calculation of the spectral integrals
is a computational task
requiring a long time,
especially if one is using the result
in another numerical integration procedure.
Therefore,
it seems reasonable to pre-compute
analytic images of couplings
for a fixed set of argument values,
consisting of $N$ points for each argument.
For example we will consider in what follows
the case of ${\mathcal A}_{\nu}^{(1);glob}[L,\nu,\Lambda_3^{(1)}]$.
We will be interested in the following ranges of arguments:
$L\in[-5,5]$,
$\Lambda^{(1)}_3\in[0.2,0.5]$,
and
$\nu=\in[0.5,1.5]$.
Then
\begin{eqnarray}
 \label{eq:11}
 \begin{array}{ccc}
  \verb'Lmin'=-5\,; & \verb'Lmax'=5\,;   & \verb'DL'=(\verb'Lmax'-\verb'Lmin')/(N-1)\,;  \\
   \nu\verb'min'=0.5\,;
                   & \nu\verb'max'=1.5\,; & \verb'D'\nu=(\nu\verb'max'-\nu\verb'min')/(N-1)\,;\\
\Lambda\verb'min'=0.2\,;
                   & \Lambda\verb'max'=0.5\,;
                                        & \verb'D'\Lambda=(\Lambda\verb'max'-\Lambda\verb'min')/(N-1)\,. \end{array}
\end{eqnarray}
\begin{figure}[b!]
 \centerline{\includegraphics[width=0.5\textwidth]{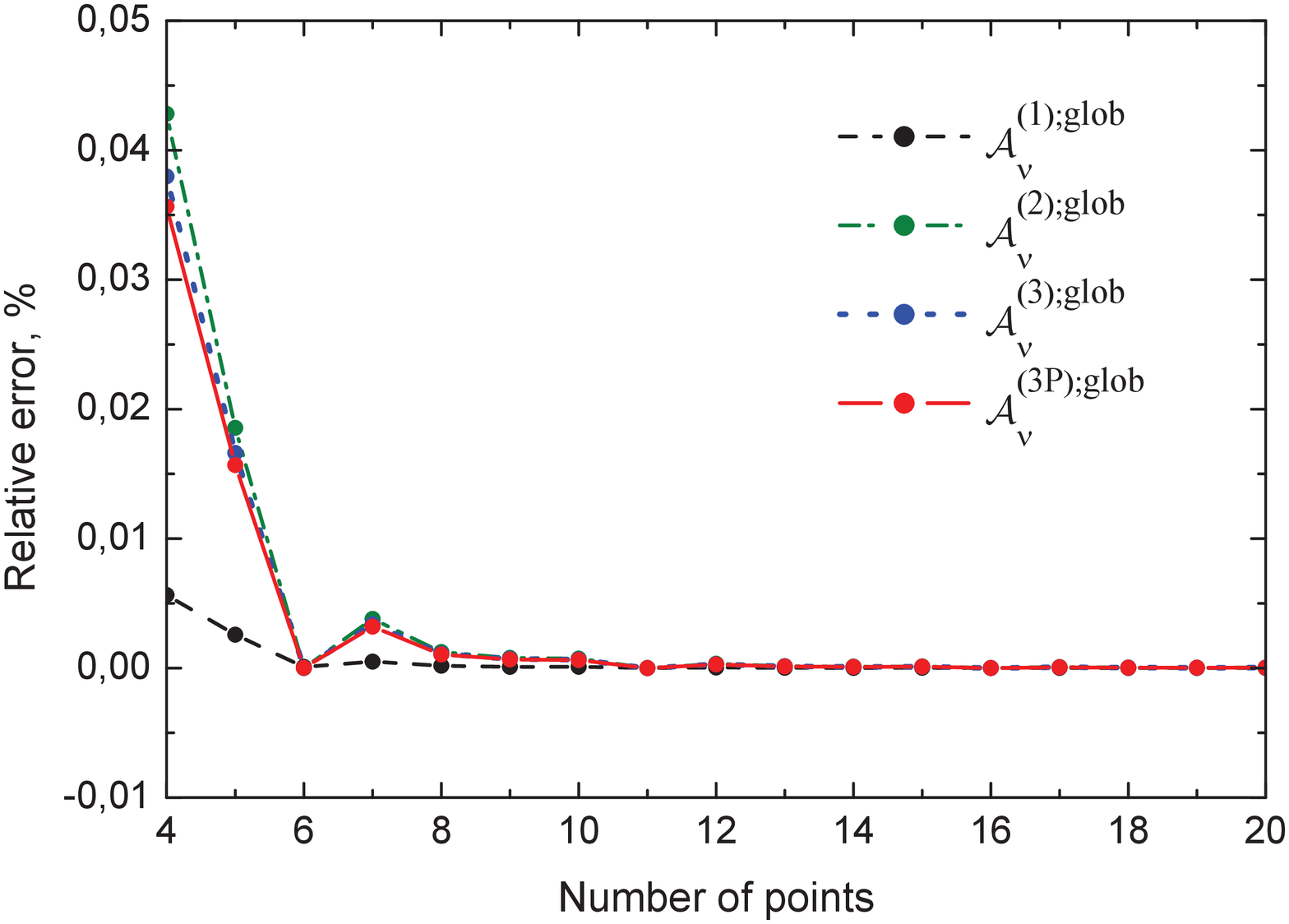}
          ~~~\includegraphics[width=0.5\textwidth]{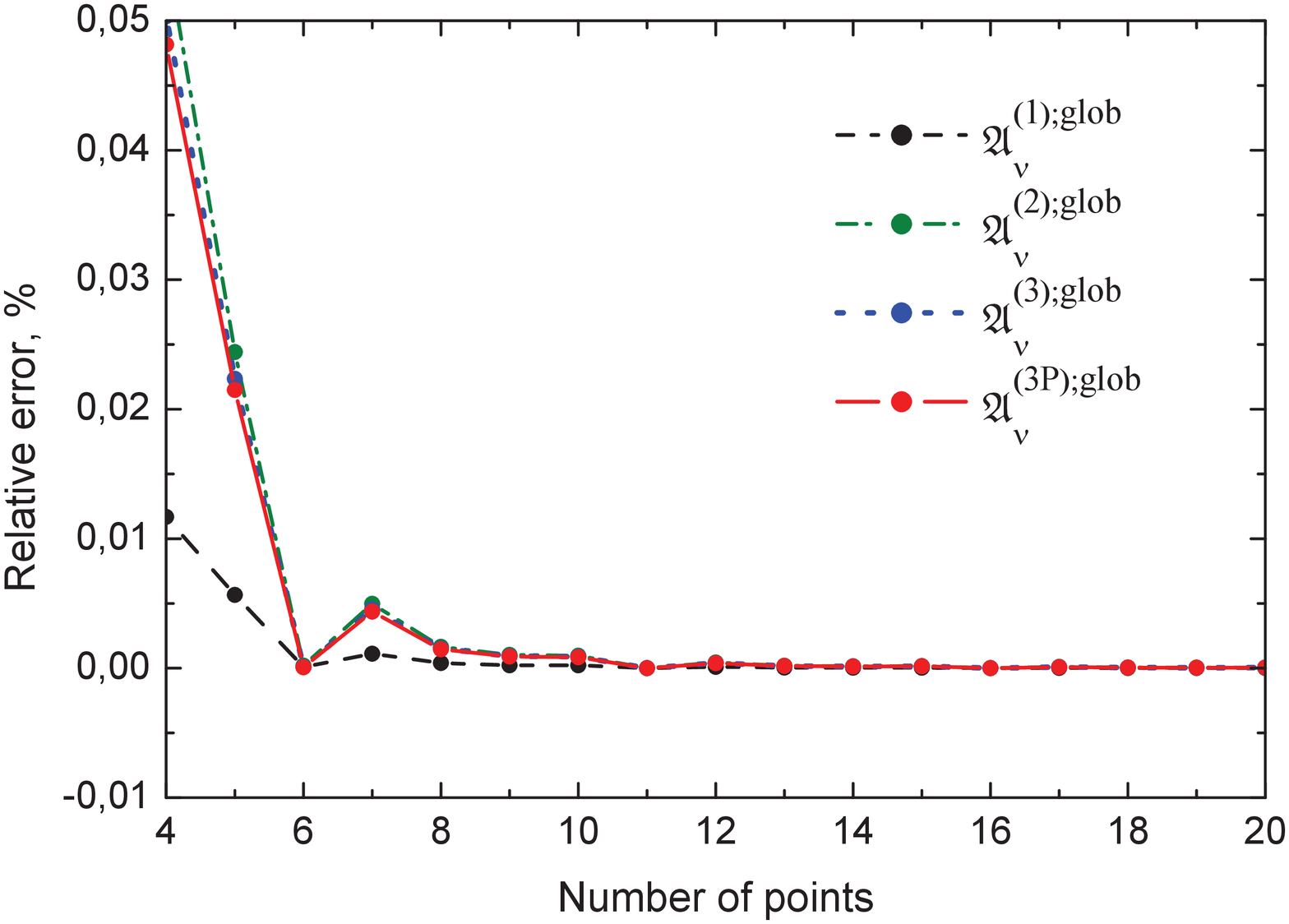}}
   \caption{\label{fig:UcalAcalN}
   \footnotesize Relative errors of the interpolation procedure for
   ${\mathcal A}_{\nu}^{(\ell);glob}$ (left panel) and
   ${\mathfrak A}_{\nu}^{(\ell);glob}$ (right panel)
   calculated at various loop orders with fixed $L=3.5$,
   $\nu=1.1$ and $\Lambda_3=0.36$ GeV.}
\end{figure}
The table of calculated values is generated by \texttt{Mathematica{ }}
using the following
command
$$
  \verb'DATA'
   = \verb'Flatten'[
      \verb'Table'[
       \verb'{'
       \verb'{'L_i,\nu_j,\Lambda_k\verb'}',
       \verb'AcalGlob1'[L_i,\nu_j,\Lambda_k]\verb'}',
       \verb'{'i,N\verb'}', \verb'{'j,N\verb'}', \verb'{'k,N\verb'}'],
       2]
$$
where $L_i=\verb'Lmin'+(i-1)\verb'DL'$,
      $\nu_j=\nu\verb'min'+(j-1)\verb'D'\nu$,
and   $\Lambda_k=\Lambda\verb'min'+(k-1)\verb'D'\Lambda$.
Then we save all calculated results in the file ``AcalGlob1i.dat'':
\begin{verbatim}
In[2] :  XY = N[DATA]; {outFile = OpenWrite["AcalGlob1i.dat"],
                       Write[outFile, XY], Close[outFile]}
Out[2]: {OutputStream["AcalGlob1i.dat", 15], Null, "AcalGlob1i.dat"}
\end{verbatim}
\begin{figure}[b!]
\centerline{\includegraphics[width=0.5\textwidth]{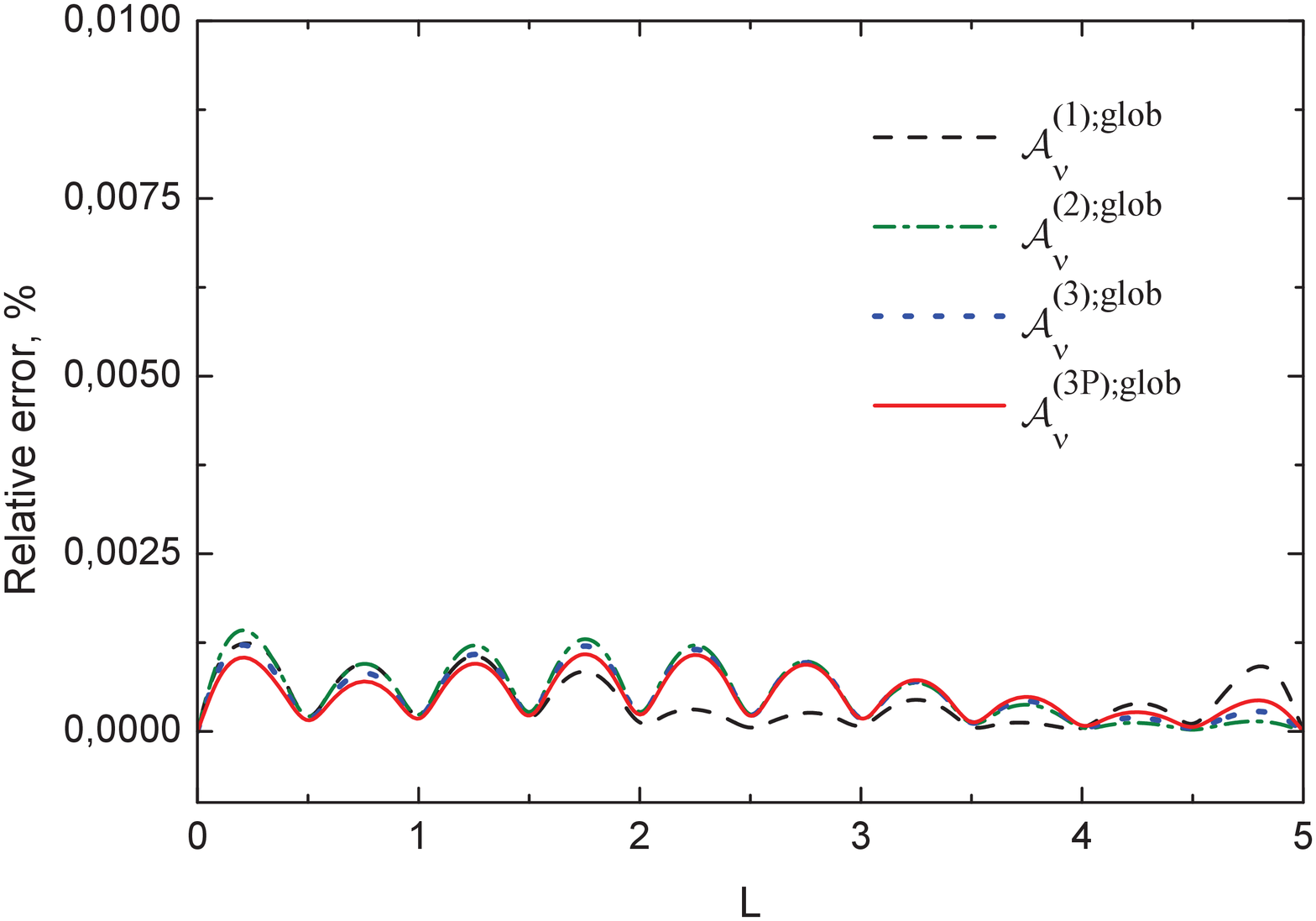}
         ~~~\includegraphics[width=0.5\textwidth]{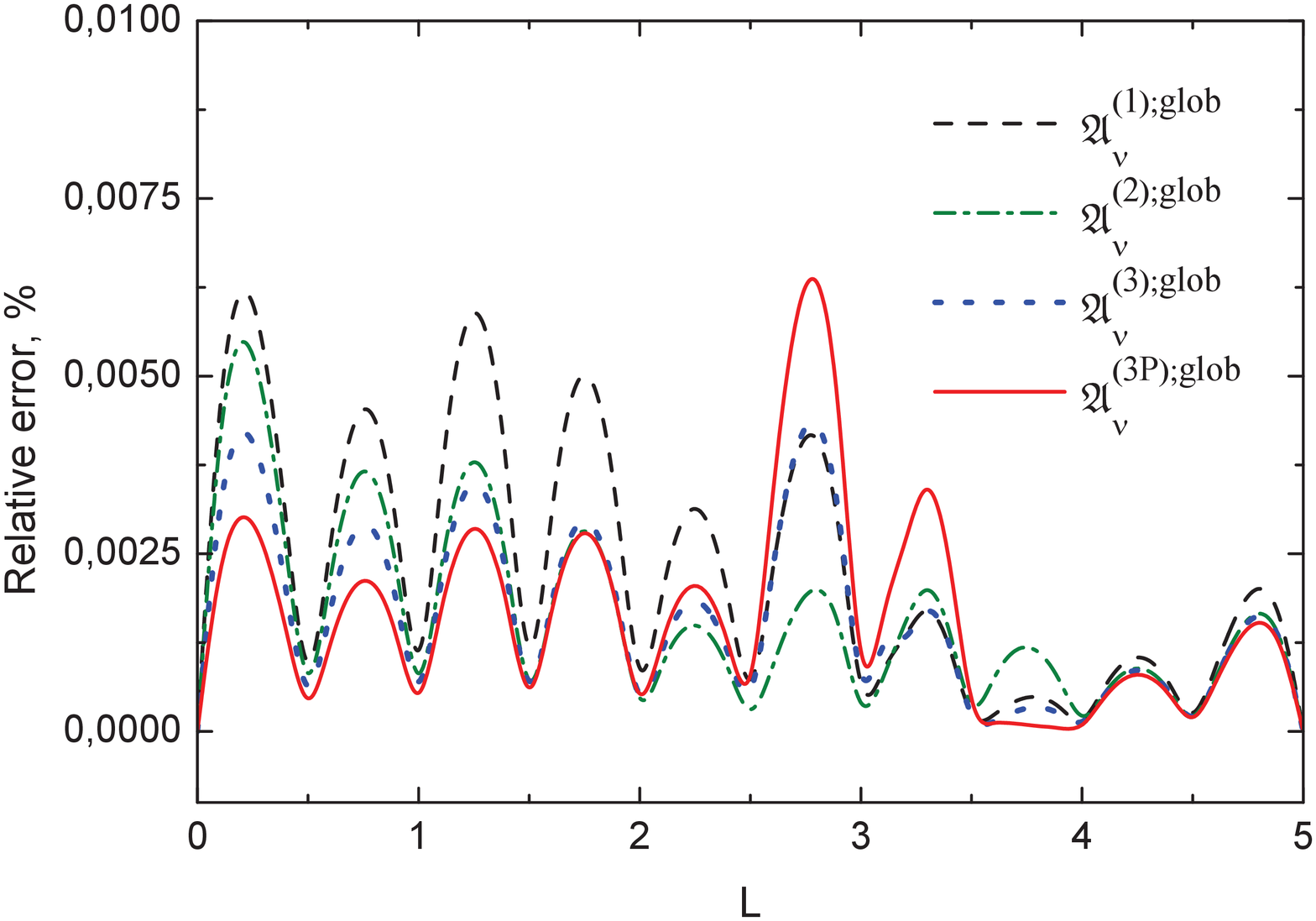}
   \vspace*{-3mm}}
   \caption{\label{fig:UcalAcalL}
   \footnotesize Relative error of the interpolation procedure
   for ${\mathcal A}_{\nu=1.1}^{glob}$ (left panel) and
   ${\mathfrak A}_{\nu=1.1}^{glob}$ (right panel), calculated
   at various loop orders with $\Lambda_3=0.36$ GeV for $N=11$
   number of points.}
\end{figure}
After that we can read them and use interpolation
to reproduce function
${\mathcal A}_{\nu}^{(1);glob}[L,\nu,\Lambda_3^{(1)}]$
in the considered ranges of arguments values:
\begin{verbatim}
In[3] :  DATA = Read["AcalGlob1i.dat"];
         AcalGlob1Interp = Interpolation[DATA]
\end{verbatim}
in order to select the appropriate value of $N$.
Now we can analyze the accuracy of inter\-polation.
In Fig~\ref{fig:UcalAcalN} we show
the dependencies of interpolation errors
on the number of the used points $N$.
One can see, that using the interpolation at $N=6$ for ${\mathcal
A}_{\nu}^{(\ell);glob}$ and ${\mathfrak A}_{\nu}^{(\ell);glob}$
provided accuracy not worse 0.005\%.

In the previous case we investigated the dependence
of the accuracy of interpolation
on the number of points at fixed $L$,
$\nu$ and $\Lambda_3$. Let us now consider how the accuracy of the
interpolation depends on the $L$. These results are shown in
Fig.~\ref{fig:UcalAcalL}.
From the last figure one can see that the maximum error of
interpolation corresponds to the region $L=0\div5$.
The error in
${\mathcal A}_{\nu=0.6}^{(1);glob}$
is less than in
${\mathfrak A}_{\nu=1.1}^{(1);glob}$.
In any case,
using $N=11$ points
for interpolation of pre-computed data
for each parameter $L$, $\nu$ and $\Lambda_3$
provides an error less than 0.01 \%.

To obtain the results much faster
one can use module \texttt{FAPT\_Interp.m}
which consists of procedures
$\text{AcalGlob}{\ell}\text{i}[L,\nu,\Lambda_3]$
and
$\text{UcalGlob}{\ell}\text{i}[L,\nu,\Lambda_3]$.
They are based on interpolation
using the basis of the precalculated data
in the ranges $L=[-5;13]$;
$\nu^\text{1-loop}=[0.5;4.0]$
and
$\Lambda^\text{1-loop}_{n_f=3}=[0.150;0.300]$;
$\nu^\text{2-loop}=[0.5;5.0]$
and
$\Lambda^\text{2-loop}_{n_f=3}=[0.300;0.450]$;
$\nu^\text{3-loop}=[0.5;6.0]$
and
$\Lambda^\text{3-loop}_{n_f=3}=[0.300;0.450]$;
$\nu^\text{4-loop}=[0.5;7.0]$
and
$\Lambda^\text{4-loop}_{n_f=3}=[0.300;0.450]$.
For example, in the four-loop case
module \texttt{FAPT\_Interp.m} contains procedures
\begin{verbatim}
AcalGlob4i = Interpolation[Read[".\\sources\\AcalGlob4i.dat"]];
UcalGlob4i = Interpolation[Read[".\\sources\\UcalGlob4i.dat"]];
\end{verbatim}
which should be used with the same arguments
$L$, $\nu$, and $\Lambda_3$
as the original procedures
$\text{AcalGlob}{\ell}[L,\nu,\Lambda_3]$
and
$\text{UcalGlob}{\ell}[L,\nu,\Lambda_3]$.
They provide much faster results of calculations
with high enough accuracy:
\begin{verbatim}
In[1]:= Timing[AcalGlob4i[1, 1.1, 0.36]]
Out[1]= {0., 0.39298}

In[2]:= Timing[AcalGlob4[1, 1.1, 0.36]]
Out[2]= {0.405, 0.392964}

In[3]:= Timing[UcalGlob4i[1, 1.1, 0.36]]
Out[3]= {0., 0.375421}

In[4]:= Timing[UcalGlob4[1, 1.1, 0.36]]
Out[4]= {0.359, 0.375372}
\end{verbatim}

\vspace*{+7mm}
\textbf{Acknowledgments}\vspace*{+1mm}

We would like to thank
Andrei Kataev, Sergey Mikhailov,
Irina Potapova,
Dmitry Shirkov, and Nico Stefanis
for stimulating discussions and useful remarks.
This work was supported in part
by the Russian Foundation for Fundamental Research
(Grant No.\ 11-01-00182)
and the BRFBR--JINR Cooperation Program
under contract No.\ F10D-002.
\vspace*{+7mm}

\begin{appendix}
\appendix
\section{Numerical parameters}
 \label{app:NumDefFAPT}
Here we shortly describe numerical parameters
used in the package.

First, in \texttt{FAPT.m} we use
the pole masses of heavy quarks and $Z$-boson,
collected in the set \texttt{NumDefFAPT}:
\begin{eqnarray}
  \begin{array}{llll}
    \texttt{MQ4}:
    &
    M_c = 1.65~\text{GeV}\,,
    &
    \texttt{MQ5}:
    &
    M_b = 4.75~\text{GeV}\,;
    \\
    \texttt{MQ6}:
    &
    M_t = 172.5~\text{GeV}\,,
    &
    \texttt{MZboson}:
    &
    M_Z = 91.19~\text{GeV}\,.
  \end{array}
  \label{eq:numdef}
\end{eqnarray}
Note here that all mass variables and parameters
are measured in GeVs.
That means, for example,
that in all procedures of our package
the following value $\texttt{MQ4}=1.65$
is used.
The package \texttt{RunDec} of~\cite{CKS00}
is using the set \texttt{NumDef}
with slightly different values of these parameters
($M_c=1.6$ GeV, $M_b=4.7$ GeV, $M_t=175$ GeV,
 $M_Z= 91.18$ GeV).

Second, we collect in the set \texttt{setbetaFAPT}
the following rules of substitutions
$b_i\to b_i(n_f)$, cf. Eq.\,(\ref{eq:beta.coef}),
\begin{eqnarray}
 \label{eq:setbeta}
   \texttt{b0}: b_0 \!&\!\to\!&\! 11 - \frac23\,n_f\,,
   \qquad
   \texttt{b1}: b_1 \to 102 - \frac{38}{3}\,n_f\,,\nonumber\\
   \texttt{b2}: b_2 \!&\!\to\!&\! \frac{2857}{2} - \frac{5033}{18}\,n_f + \frac{325}{54}\,n_f^2\,,\\
   \texttt{b3}: b_3 \!&\!\to\!&\! \frac{149753}{6}
                                - \frac{1078361}{162}\,n_f
                                + \frac{50065}{162}\,n_f^2
                                + \frac{1093}{729}\,n_f^3
   \nonumber\\
                    \!&\! \!&\! + \left[3564 - \frac{6508}{27}\,n_f
                                      + \frac{6472}{81}\,n_f^2
                                  \right]\,\zeta[3]\,.
 \nonumber
\end{eqnarray}
Here we follow the same substitution strategy as in~\cite{CKS00},
but our $b_i$ differ from theirs $b_i^\text{CKS}$ by factors $4^{i+1}$:
$b_i=4^{i+1}\,b_i^\text{CKS}$.
In parallel,
the set \texttt{setbetaFAPT4Pi}
defines substitutions $b_i\to b_i(n_f)/(4\pi)$
which are more appropriate
to determine coefficients $c_i(n_f)$.

\vspace*{7mm}
\section{Description of the main procedures}
\label{sec:def}
Here we shortly describe the main procedures of our package
which can be useful for practical calculations.

\begin{flushleft}
\begin{itemize}
 \item \texttt{RhoBar}$\ell$\texttt{[L,Nf,Nu]}:
  \begin{itemize}
  \item[\textit{general:}] it computes the $\ell$-loop spectral density
   $\bar{\rho}^{(\ell)}[L_\sigma, n_f, \nu]$;
  \item[\textit{input:}] the logarithmic argument \texttt{L}=$L_\sigma=\ln[\sigma/\Lambda^2]$,
   the number of active flavors \texttt{Nf}=$n_f$,
   and the power index \texttt{Nu}=$\nu$;
  \item[\textit{output:}] $\bar{\rho}^{(\ell)}$;
  \item[\textit{example:}] In order to compute the value of
    the four-loop spectral density
    $\bar{\rho}^{(4)}[3.95, 4, 1.62]=0.0247209$
    one has to use the command
    \verb|RhoBar4[3.95, 4, 1.62]|.
  \end{itemize}

 \item \texttt{RhoGlob}$\ell$\texttt{[L,Nu,Lam]}:
  \begin{itemize}
  \item[\textit{general:}] it computes the $\ell$-loop global spectral density
   $\rho^{(\ell);\text{glob}}[L_\sigma, \nu, \Lambda_{n_f=3}]$;
  \item[\textit{input:}] the logarithmic argument \texttt{L}=$L_\sigma=\ln[\sigma/\Lambda_{n_f=3}^2]$,
   the power index \texttt{Nu}=$\nu$,
   and the QCD scale parameter \texttt{Lam}=$\Lambda_{n_f=3}$ (in GeV);
  \item[\textit{output:}] $\rho^{(\ell);\text{glob}}$;
  \item[\textit{example:}] In order to compute the value of
    the four-loop spectral density
    $\rho^{(4);\text{glob}}[3.95, 1.62, 0.350]=0.0221662$
    one has to use the command
    \verb|RhoGlob4[3.95, 1.62, 0.35]|.
  \end{itemize}
\newpage

 \item \texttt{AcalBar}$\ell$\texttt{[L,Nf,Nu]}:
  \begin{itemize}
  \item[\textit{general:}] it computes the $\ell$-loop $n_f$-fixed
   analytic coupling $\bar{\mathcal A}_\nu^{(\ell)}[L, n_f]$
   in Euclidean domain;
  \item[\textit{input:}] the logarithmic argument \texttt{L}=$\ln[Q^2/\Lambda^2]$,
   the number of active flavors \texttt{Nf}=$n_f$,
   and the power index \texttt{Nu}=$\nu$;
  \item[\textit{output:}] $\bar{\mathcal A}_\nu^{(\ell)}$;
  \item[\textit{example:}] In order to compute the value of
    the three-loop spectral density
    $\bar{\mathcal A}_{1.62}^{(3)}[3.95, 4]=0.11352$
    one has to use the command
    \verb|AcalBar3[3.95, 4, 1.62]|.
  \end{itemize}

 \item \texttt{UcalBar}$\ell$\texttt{[L,Nf,Nu]}:
  \begin{itemize}
  \item[\textit{general:}] it computes the $\ell$-loop $n_f$-fixed
   analytic coupling $\bar{\mathfrak A}_\nu^{(\ell)}[L, n_f]$
   in Minkowski domain;
  \item[\textit{input:}] the logarithmic argument \texttt{L}=$\ln[s/\Lambda^2]$,
   the number of active flavors \texttt{Nf}=$n_f$,
   and the power index \texttt{Nu}=$\nu$;
  \item[\textit{output:}] $\bar{\mathfrak A}_\nu^{(\ell)}$;
  \item[\textit{example:}] In order to compute the value of
    the three-loop spectral density
    $\bar{\mathfrak A}_{1.62}^{(3)}[3.95, 4]=0.1011$
    one has to use the command
    \verb|UcalBar3[3.95, 4, 1.62]|.
  \end{itemize}

 \item \texttt{AcalGlob}$\ell$\texttt{[L,Nu,Lam]}:
  \begin{itemize}
  \item[\textit{general:}] it computes the $\ell$-loop global analytic coupling
   $\mathcal A_\nu^{(\ell);\text{glob}}[L, \nu, \Lambda_{n_f=3}]$
   in Euclidean domain;
  \item[\textit{input:}] the logarithmic argument \texttt{L}=$L_\sigma=\ln[\sigma/\Lambda_{n_f=3}^2]$,
   the power index \texttt{Nu}=$\nu$,
   and the QCD scale parameter \texttt{Lam}=$\Lambda_{n_f=3}$ (in GeV);
  \item[\textit{output:}] $\mathcal A_\nu^{(\ell);\text{glob}}$;
  \item[\textit{example:}] In order to compute the value of
    the two-loop analytic coupling
    $\mathcal A_{1.62}^{(2);\text{glob}}[3.95, 0.350]=0.103858$
    one has to use the command
    \verb|AcalGlob2[3.95, 1.62, 0.35]|.
  \end{itemize}

 \item \texttt{UcalGlob}$\ell$\texttt{[L,Nu,Lam]}:
  \begin{itemize}
  \item[\textit{general:}] it computes the $\ell$-loop global analytic coupling
   $\mathfrak A_\nu^{(\ell);\text{glob}}[L, \nu, \Lambda_{n_f=3}]$
   in Minkowski domain;
  \item[\textit{input:}] the logarithmic argument \texttt{L}=$\ln[s/\Lambda_{n_f=3}^2]$,
   the power index \texttt{Nu}=$\nu$,
   and the QCD scale parameter \texttt{Lam}=$\Lambda_{n_f=3}$ (in GeV);
  \item[\textit{output:}] $\mathfrak A_\nu^{(\ell);\text{glob}}$;
  \item[\textit{example:}] In order to compute the value of
    the two-loop analytic coupling
    $\mathfrak A_{1.62}^{(2);\text{glob}}[3.95, 0.350]=0.0932096$
    one has to use the command
    \verb|UcalGlob2[3.95, 1.62, 0.35]|.
  \end{itemize}
\end{itemize}
\end{flushleft}

All $\Lambda_{n_f=3}$ are in GeV,
all squared momentum transfer $Q^2$ (Euclidean),
central-of-mass energy squared $s$ (Minkowski),
and spectral-integration variables $\sigma$ are in GeV$^2$.
The number of loops $\ell$ is everywhere specified
in the name of a procedure.

\end{appendix}


\begin{thebibliography}{10}
\expandafter\ifx\csname url\endcsname\relax
  \def\url#1{}\fi
\expandafter\ifx\csname urlprefix\endcsname\relax\def\urlprefix{URL }\fi
\expandafter\ifx\csname href\endcsname\relax
  \def\href#1#2{} \def\path#1{}\fi


\bibitem{SS}
 D.~V. Shirkov, I.~L. Solovtsov,
  \textit{Analytic QCD running coupling with finite IR behaviour and universal
   $\bar{\alpha}_s(0)$ value},
  JINR Rapid Commun. 2[76] (1996) 5--10.
  \newblock \href {http://arxiv.org/abs/hep-ph/9604363}
  {\path{arXiv:hep-ph/9604363}}\\
  \textit{Analytic model for the QCD running coupling with universal
   $\bar{\alpha}_s(0)$ value},
  Phys. Rev. Lett. 79 (1997) 1209--1212.
  \newblock \href {http://arxiv.org/abs/hep-ph/9704333}
  {\path{arXiv:hep-ph/9704333}}

\bibitem{MS96}
 K.~A. Milton, I.~L. Solovtsov,
  \textit{Analytic perturbation theory in QCD and Schwinger's connection
   between the beta function and the spectral density},
  Phys. Rev. D55 (1997) 5295--5298.
  \newblock \href {http://arxiv.org/abs/hep-ph/9611438}
  {\path{arXiv:hep-ph/9611438}}

\bibitem{SS98}
 I.~L. Solovtsov, D.~V. Shirkov,
  \textit{Analytic approach to perturbative QCD and renormalization
   scheme dependence},
  Phys. Lett. B442 (1998) 344--348.
  \newblock \href {http://arxiv.org/abs/hep-ph/9711251}
  {\path{arXiv:hep-ph/9711251}}

\bibitem{BMS05}
 A.~P. Bakulev, S.~V. Mikhailov, N.~G. Stefanis,
  \textit{QCD analytic perturbation theory: From integer powers
   to any power of the running coupling},
  Phys. Rev. D72 (2005) 074014;
  Erratum: \textit{ibid.} D72 (2005) 119908(E).
  \newblock \href {http://arxiv.org/abs/hep-ph/0506311}
  {\path{arXiv:hep-ph/0506311}}

\bibitem{BKS05}
 A.~P. Bakulev, A.~I. Karanikas, N.~G. Stefanis,
  \textit{Analyticity properties of three-point functions in QCD
  beyond leading order},
  Phys. Rev. D72 (2005) 074015.
  \newblock \href {http://arxiv.org/abs/hep-ph/0504275}
  {\path{arXiv:hep-ph/0504275}}

\bibitem{BMS06}
 A.~P. Bakulev, S.~V. Mikhailov, N.~G. Stefanis,
  \textit{Fractional analytic perturbation theory in Minkowski
   space and application to Higgs boson decay into a
   $b\bar{b}$ pair},
  Phys. Rev. D75 (2007) 056005;
  Erratum: \textit{ibid.} D77 (2008) 079901(E).
  \newblock \href {http://arxiv.org/abs/hep-ph/0607040}
  {\path{arXiv:hep-ph/0607040}}

\bibitem{LAH56}
 L.~D. Landau, A.~Abrikosov, L.~Halatnikov,
  \textit{On the quantum theory of fields},
  Nuovo Cim. Suppl. 3 (1956) 80--104.

\bibitem{DG05}
 D.~J. Gross,
  \textit{The discovery of asymptotic freedom and the emergence of {QCD}},
  Proc. Nat. Acad. Sci. 102 (2005) 9099--9108.

\bibitem{BS59}
 N.~N. Bogolyubov, D.~V. Shirkov,
  \textit{Introduction to the theory of quantized fields},
  Intersci. Monogr. Phys. Astron. 3 (1959) 1--720.

\bibitem{BSvvtkp}
 N.~N. Bogolyubov, D.~V. Shirkov,
  \textit{Introduction to the Theory of Quantum Fields},
   Wiley, New York, 1959, 1980.

\bibitem{BLT69}
 N.~N. Bogolyubov, A.~A. Logunov, I.~T. Todorov,
  \textit{Introduction to Axiomatic Quantum Field Theory},
   Benjamin Cummings, Reading, Massachusetts, 1975.

\bibitem{BLS60}
 N.~N. Bogolyubov, A.~A. Logunov, D.~V. Shirkov,
  \textit{The method of dispersion relations and perturbation theory},
  Soviet Physics JETP 10 (1960) 574.

\bibitem{Rad82}
 A.~V. Radyushkin,
  \textit{Optimized lambda-parametrization for the QCD running  coupling constant in space-like and time-like regions},
  JINR Rapid Commun. 78 (1996) 96--99,
  [JINR Preprint, E2-82-159, 26 Febr. 1982].
  \newblock \href {http://arxiv.org/abs/hep-ph/9907228}
  {\path{arXiv:hep-ph/9907228}}

\bibitem{KP82}
 N.~V. Krasnikov, A.~A. Pivovarov,
  \textit{The influence of the analytical continuation effects on the value of the QCD scale parameter $\Lambda$
   extracted from the data on charmonium and upsilon hadron decays},
  Phys. Lett. B116 (1982) 168--170.

\bibitem{JS95-349}
 H.~F. Jones, I.~L. Solovtsov,
  \textit{QCD running coupling constant in the timelike region},
  Phys. Lett. B349 (1995) 519--524.
  \newblock \href {http://arxiv.org/abs/hep-ph/9501344}
  {\path{arXiv:hep-ph/9501344}}

\bibitem{BB95}
 M.~Beneke, V.~M. Braun,
  \textit{Naive non-Abelianization and resummation of fermion bubble chains},
  Phys. Lett. B348 (1995) 513--520.
  \newblock \href {http://arxiv.org/abs/hep-ph/9411229}
  {\path{arXiv:hep-ph/9411229}}

\bibitem{BBB95}
 P.~Ball, M.~Beneke, V.~M. Braun,
  \textit{Resummation of $(\beta_0\alpha_s)^n$ corrections in QCD:
   Techniques and applications to the tau hadronic width and
   the heavy quark pole mass},
   Nucl. Phys. B452 (1995) 563--625.
   \newblock \href {http://arxiv.org/abs/hep-ph/9502300}
   {\path{arXiv:hep-ph/9502300}}

\bibitem{Mag99}
 B.~A. Magradze,
  \textit{Analytic approach to perturbative QCD},
  Int. J. Mod. Phys. A15 (2000) 2715--2734.
  \newblock \href {http://arxiv.org/abs/hep-ph/9911456}
  {\path{arXiv:hep-ph/9911456}}

\bibitem{Kour99}
 D.~S. Kourashev,
  \textit{The QCD observables expansion over the scheme-independent
   two-loop coupling constant powers, the scheme dependence reduction},
  hep-ph/9912410 (1999).
  \newblock \href {http://arxiv.org/abs/hep-ph/9912410}
  {\path{arXiv:hep-ph/9912410}}

\bibitem{Mag00}
 B.~A. Magradze,
  \textit{QCD coupling up to third order in standard and analytic
   perturbation theories},
  \uppercase{D}ubna preprint E2-2000-222, 2000
  [hep-ph/0010070].
  \newblock \href {http://arxiv.org/abs/hep-ph/0010070}
  {\path{arXiv:hep-ph/0010070}}

\bibitem{KM01}
 D.~S. Kourashev, B.~A. Magradze,
  \textit{Explicit expressions for Euclidean and Min\-kowskian QCD observables
   in analytic perturbation theory},
  \uppercase{P}reprint RMI-2001-18, 2001 [hep-ph/0104142] (2001).
  \newblock \href {http://arxiv.org/abs/hep-ph/0104142}
  {\path{arXiv:hep-ph/0104142}}

\bibitem{Mag03u}
 B.~A. Magradze,
  \textit{Practical techniques of analytic perturbation theory of QCD},
  \uppercase{P}reprint RMI-2003-55, 2003 [hep-ph/0305020] (2003).
  \newblock \href {http://arxiv.org/abs/hep-ph/0305020}
  {\path{arXiv:hep-ph/0305020}}

\bibitem{KM03}
 D.~S. Kourashev, B.~A. Magradze,
 \textit{Explicit expressions for timelike and spacelike observables
  of quantum chromodynamics in analytic perturbation theory},
  Theor. Math. Phys. 135 (2003) 531--540.

\bibitem{Mag05}
 B.~A. Magradze,
  \textit{A novel series solution to the renormalization group equation in QCD},
  Few Body Syst. 40 (2006) 71--99.
  \newblock \href {http://arxiv.org/abs/hep-ph/0512374}
  {\path{arXiv:hep-ph/0512374}}

\bibitem{MSSY00}
 K.~A. Milton, I.~L. Solovtsov, O.~P. Solovtsova, V.~I. Yasnov,
  \textit{Renormalization scheme and higher loop stability in hadronic tau decay
   within analytic perturbation theory},
  Eur. Phys. J. C14 (2000) 495--501.
  \newblock \href {http://arxiv.org/abs/hep-ph/0003030}
  {\path{arXiv:hep-ph/0003030}}

\bibitem{MSS01}
 K.~A. Milton, I.~L. Solovtsov, O.~P. Solovtsova,
  \textit{Remark on the perturbative component of inclusive tau decay},
  Phys. Rev. D65 (2002) 076009.
  \newblock \href {http://arxiv.org/abs/hep-ph/0111197}
  {\path{arXiv:hep-ph/0111197}}

\bibitem{CvVa06}
 G.~Cvetic, C.~Valenzuela,
  \textit{Various versions of analytic QCD and skeleton-motivated evaluation of observables},
  Phys. Rev. D74 (2006) 114030.
  \newblock \href {http://arxiv.org/abs/hep-ph/0608256}
  {\path{arXiv:hep-ph/0608256}}

\bibitem{CKV09}
 G.~Cvetic, R.~Kogerler, C.~Valenzuela,
  \textit{Analytic QCD coupling with no power terms in UV regime},
  J. Phys. G37 (2010) 075001.
  \newblock \href {http://arxiv.org/abs/0912.2466}
  {\path{arXiv:0912.2466}}

\bibitem{Mag10}
 B.~A. Magradze,
  \textit{Testing the Concept of Quark-Hadron Duality with the ALEPH
   $\tau$ Decay Data},
  Few Body Syst. 48 (2010) 143--169.
  \newblock \href {http://arxiv.org/abs/1005.2674}
  {\path{arXiv:1005.2674}}

\bibitem{MSS98}
 K.~A. Milton, I.~L. Solovtsov, O.~P. Solovtsova,
  \textit{The Bjorken sum rule in the analytic approach to perturbative QCD},
  Phys. Lett. B439 (1998) 421--427.
  \newblock \href {http://arxiv.org/abs/hep-ph/9809510}
  {\path{arXiv:hep-ph/9809510}}

\bibitem{PST08}
 R.~S. Pasechnik, D.~V. Shirkov, O.~V. Teryaev,
  \textit{Bjorken Sum Rule and pQCD frontier on the move},
  Phys. Rev. D78 (2008) 071902.
  \newblock \href {http://arxiv.org/abs/0808.0066}
  {\path{arXiv:0808.0066}}

\bibitem{MSS98GLS}
 K.~A. Milton, I.~L. Solovtsov, O.~P. Solovtsova,
  \textit{The Gross--Llewellyn Smith sum rule in the analytic approach to perturbative QCD},
  Phys. Rev. D60 (1999) 016001.
  \newblock \href {http://arxiv.org/abs/hep-ph/9809513}
  {\path{arXiv:hep-ph/9809513}}

\bibitem{SZ05}
 D.~V. Shirkov, A.~V. Zayakin,
  \textit{Analytic perturbation theory for practitioners and Upsilon decay},
  Phys. Atom. Nucl. 70 (2007) 775--783.
  \newblock \href {http://arxiv.org/abs/hep-ph/0512325}
  {\path{arXiv:hep-ph/0512325}}

\bibitem{SSK99}
 N.~G. Stefanis, W.~Schroers, H.-C. Kim,
  \textit{Pion form factors with improved infrared factorization},
  Phys. Lett. B449 (1999) 299.
  \newblock \href {http://arxiv.org/abs/hep-ph/9807298}
  {\path{arXiv:hep-ph/9807298}}

\bibitem{SSK00}
 N.~G. Stefanis, W.~Schroers, H.-C. Kim,
  \textit{Analytic coupling and Sudakov effects in exclusive processes:
  Pion and $\gamma^*\gamma\to\pi^0$ form factors},
  Eur. Phys. J. C18 (2000) 137--156.
  \newblock \href {http://arxiv.org/abs/hep-ph/0005218}
  {\path{arXiv:hep-ph/0005218}}

\bibitem{BPSS04}
 A.~P. Bakulev, K.~Passek-Kumeri\v{c}ki, W.~Schroers, N.~G. Stefanis,
  \textit{Pion form factor in QCD: From nonlocal condensates to {NLO} analytic perturbation theory},
  Phys. Rev. D70 (2004) 033014.
  \newblock \href {http://arxiv.org/abs/hep-ph/0405062}
  {\path{arXiv:hep-ph/0405062}}

\bibitem{SS99}
 I.~L. Solovtsov, D.~V. Shirkov,
  \textit{The analytic approach in quantum chromodynamics},
  Theor. Math. Phys. 120 (1999) 1220--1244.
  \newblock \href {http://arxiv.org/abs/hep-ph/9909305}
  {\path{arXiv:hep-ph/9909305}}

\bibitem{Shi00}
 D.~V. Shirkov,
  \textit{Analytic perturbation theory for QCD observables},
  Theor. Math. Phys. 127 (2001) 409--423.
  \newblock \href {http://arxiv.org/abs/hep-ph/0012283}
  {\path{arXiv:hep-ph/0012283}}

\bibitem{SS06}
 D.~V. Shirkov, I.~L. Solovtsov,
  \textit{Ten years of the analytic perturbation theory in QCD},
  Theor. Math. Phys. 150 (2007) 132--152.
  \newblock \href {http://arxiv.org/abs/hep-ph/0611229}
  {\path{arXiv:hep-ph/0611229}}

\bibitem{DPTar89}
 L.~V. Dung, H.~D. Phuoc, O.~V. Tarasov,
  \textit{The influence of quark masses on the infrared behavior of $\alpha_s(Q^2)$ in QCD},
  Sov. J. Nucl. Phys. 50 (1989) 1072--1079.

\bibitem{Sim01}
 Y.~A. Simonov,
  \textit{Perturbative expansions in QCD and analytic properties of $\alpha_s$},
  Phys. Atom. Nucl. 65 (2002) 135--152.
  \newblock \href {http://arxiv.org/abs/hep-ph/0109081}
  {\path{arXiv:hep-ph/0109081}}

\bibitem{NP04}
 A.~V. Nesterenko, J.~Papavassiliou,
  \textit{The massive analytic invariant charge in QCD},
  Phys. Rev. D71 (2005) 016009.
  \newblock \href {http://arxiv.org/abs/hep-ph/0410406}
  {\path{arXiv:hep-ph/0410406}}

\bibitem{BaSh11}
 A.~P. Bakulev, D.~V. Shirkov,
  {Inevitability and Importance of Non-Perturbative Elements in Quantum Field Theory},
  in: B.~Dragovich, Z.~Raki$\grave{\text{c}}$ (Eds.),
  Proceedings of the 6th MATHEMATICAL PHYSICS MEETING:
  Summer School and Conference on Modern Mathematical Physics,
  September 14--23, 2010, Belgrade, Serbia, Institute of Physics,
  Belgrade (Serbia), 2011, pp. 27--53.
  \newblock \href {http://arxiv.org/abs/arXiv:1102.2380}
  {\path{arXiv:arXiv:1102.2380}}

\bibitem{Shi12}
 D.~V. Shirkov,
  {A Few Lessons from pQCD Analysis at Low Energies},
  arXiv:1202.3220 [hep-ph] (2012).
  \newblock \href {http://arxiv.org/abs/1202.3220}
  {\path{arXiv:1202.3220}}

\bibitem{KS01}
 A.~I. Karanikas, N.~G. Stefanis,
  \textit{Analyticity and power corrections in hard-scattering hadronic functions},
  Phys. Lett. B504 (2001) 225--234;
  Erratum: \textit{ibid.} B636 (2006) 330.
  \newblock \href {http://arxiv.org/abs/hep-ph/0101031}
  {\path{arXiv:hep-ph/0101031}}

\bibitem{Ste02}
 N.~G. Stefanis,
  \textit{Perturbative logarithms and power corrections in QCD hadronic functions:
   A unifying approach},
  Lect. Notes Phys. 616 (2003) 153--166.
  \newblock \href {http://arxiv.org/abs/hep-ph/0203103}
  {\path{arXiv:hep-ph/0203103}}

\bibitem{AB08}
 A.~P. Bakulev,
  \textit{Global Fractional Analytic Perturbation Theory in QCD
   with Selected Applications},
  Phys. Part. Nucl. 40 (2009) 715--756.
  \newblock \href {http://arxiv.org/abs/arXiv:0805.0829 [hep-ph]}
  {\path{arXiv:0805.0829 [hep-ph]}}

\bibitem{CvKo11fapt}
 G.~Cvetic, A.~V. Kotikov,
  \textit{Analogs of noninteger powers in general analytic QCD},
  J. Phys. G39 (2012) 065005.
  \newblock \href {http://arxiv.org/abs/1106.4275}
  {\path{arXiv:1106.4275}}

\bibitem{BKM01}
 D.~J. Broadhurst, A.~L. Kataev, C.~J. Maxwell,
  {Renormalons and multiloop estimates in scalar correlators,
   Higgs decay and quark-mass sum rule},
   Nucl. Phys. B592 (2001) 247--293.
  \newblock \href {http://arxiv.org/abs/hep-ph/0007152}
  {\path{arXiv:hep-ph/0007152}}

\bibitem{BMS10}
 A.~P. Bakulev, S.~V. Mikhailov, N.~G. Stefanis,
  \textit{Higher-order QCD perturbation theory in different schemes: From FOPT to CIPT to FAPT},
  JHEP 1006 (2010) 085 (1--38).
  \newblock \href {http://arxiv.org/abs/1004.4125}
  {\path{arXiv:1004.4125}}

\bibitem{PSTSK09}
 R.~S. Pasechnik, D.~V. Shirkov, O.~V. Teryaev, O.~P. Solovtsova, V.~L. Khandramai,
  \textit{Nucleon spin structure and pQCD frontier on the move},
  Phys. Rev. D81 (2010) 016010.
  \newblock \href {http://arxiv.org/abs/0911.3297}
  {\path{arXiv:0911.3297}}

\bibitem{CIKK09}
 G.~Cvetic, A.~Y. Illarionov, B.~A. Kniehl, A.~V. Kotikov,
  \textit{Small-$x$ behavior of the structure function $F_2$ and its slope
  $\partial \ln(F_2)/\partial\ln(1/x)$ for 'frozen' and analytic strong-coupling constants},
  Phys. Lett. B679 (2009) 350--354.
  \newblock \href {http://arxiv.org/abs/0906.1925}
  {\path{arXiv:0906.1925}}

\bibitem{KotKri10}
 A.~V. Kotikov, V.~G. Krivokhizhin, B.~G. Shaikhatdenov,
  \textit{Analytic and 'frozen' QCD coupling constants up to NNLO from DIS data},
   Phys. Atom. Nucl. 75 (2012) 507--524.
   \newblock \href {http://arxiv.org/abs/1008.0545}
   {\path{arXiv:1008.0545}}

\bibitem{NeSi10}
 A.~V. Nesterenko, C.~Simolo,
  \textit{QCDMAPT: program package for Analytic approach to QCD},
  Comput. Phys. Commun. 181 (2010) 1769--1775.
  \newblock \href {http://arxiv.org/abs/1001.0901}
  {\path{arXiv:1001.0901}}


\bibitem{NeSi11}
 A.~Nesterenko, C.~Simolo,
  \textit{QCDMAPT\_F: Fortran version of QCDMAPT package},
  Comput. Phys. Commun. 182 (2011) 2303--2304.
  \newblock \href {http://arxiv.org/abs/1107.1045}
  {\path{arXiv:1107.1045}}

\bibitem{math88}
 S.~Wolfram,
  \textit{Mathematica --- a system for doing mathematics by computer},
   Addison-Wesley, New York, 1988.

\bibitem{GW73}
 D.~J. Gross, F.~Wilczek,
  \textit{Ultraviolet behavior of nonabelian gauge theories},
  Phys. Rev. Lett. 30 (1973) 1343--1346.

\bibitem{GW73a}
 D.~J. Gross, F.~Wilczek,
 \textit{Asymptotically free gauge theories. 1},
  Phys. Rev. D8 (1973) 3633--3652.

\bibitem{Pol73}
 H.~D. Politzer,
  \textit{Reliable perturbative results for strong interactions?},
  Phys. Rev. Lett. 30 (1973) 1346--1349.

\bibitem{Jon74}
 D.~R.~T. Jones,
  \textit{Two-Loop Diagrams in Yang--Mills Theory},
  Nucl. Phys. B75 (1974) 531.
  \newblock \href {http://dx.doi.org/10.1016/0550-3213(74)90093-5}
  {\path{doi:10.1016/0550-3213(74)90093-5}}

\bibitem{Cas74}
 W.~E. Caswell,
  \textit{Asymptotic Behavior of Non-Abelian Gauge Theories to Two-Loop Order},
  Phys. Rev. Lett. 33 (1974) 244.
  \newblock \href {http://dx.doi.org/10.1103/PhysRevLett.33.244}
  {\path{doi:10.1103/PhysRevLett.33.244}}

\bibitem{EgTar78}
 E.~Egorian, O.~V. Tarasov,
  \textit{Two-loop renormalization of the QCD in an arbitrary gauge},
  Theor. Math. Phys. 41 (1979) 863-- 869.

\bibitem{TVZ80}
 O.~V. Tarasov, A.~A. Vladimirov, A.~Y. Zharkov,
  \textit{The Gell-Mann--Low Function of QCD in the Three-Loop Approximation},
  Phys. Lett. B93 (1980) 429--432.
  \newblock \href {http://dx.doi.org/10.1016/0370-2693(80)90358-5}
  {\path{doi:10.1016/0370-2693(80)90358-5}}

\bibitem{LaVe93}
 S.~A. Larin, J.~A.~M. Vermaseren,
  \textit{The three-loop QCD beta function and anomalous dimensions},
  Phys. Lett. B303 (1993) 334--336.
  \newblock \href {http://arxiv.org/abs/hep-ph/9302208}
  {\path{arXiv:hep-ph/9302208}}
   \href{http://dx.doi.org/10.1016/0370-2693(93)91441-O}
        {\path{doi:10.1016/0370-2693(93)91441-O}}

\bibitem{vRVL97}
 T.~van Ritbergen, J.~A.~M. Vermaseren, S.~A. Larin,
  \textit{The four-loop beta function in quantum chromodynamics},
  Phys. Lett. B400 (1997) 379--384.
  \newblock \href {http://arxiv.org/abs/hep-ph/9701390}
  {\path{arXiv:hep-ph/9701390}}

\bibitem{Cz04}
 M.~Czakon,
  \textit{The Four-loop QCD beta-function and anomalous dimensions},
  Nucl. Phys. B710 (2005) 485--498.
  \newblock \href {http://arxiv.org/abs/hep-ph/0411261}
  {\path{arXiv:hep-ph/0411261}}

\bibitem{Mag98}
 B.~A. Magradze,
  \textit{The gluon propagator in analytic perturbation theory},
  in: F.~L. Bezrukov, V.~A. Matveev, V.~A. Rubakov, A.~N. Tavkhelidze,
  S.~V. Troitsky (Eds.),
  Proceedings of the 10th International Seminar {Quarks'98},
  Suzdal, Russia, 18--24 May 1998,
  INR RAS, Moscow, 1999, pp. 158--171.
  \newblock \href {http://arxiv.org/abs/hep-ph/9808247}
  {\path{arXiv:hep-ph/9808247}}

\bibitem{GGK98}
 E.~Gardi, G.~Grunberg, M.~Karliner,
  \textit{Can the QCD running coupling have a causal analyticity structure?},
  JHEP 07 (1998) 007.
  \newblock \href {http://arxiv.org/abs/hep-ph/9806462}
  {\path{arXiv:hep-ph/9806462}}

\bibitem{GaKa11}
 A.~V. Garkusha, A.~L. Kataev,
  {The absence of QCD $\beta$-function factorization property
   of the generalized Crewther relation in the 't Hooft $\bar{MS}$-based scheme},
   Phys. Lett. B705 (2011) 400--404.
  \newblock \href {http://arxiv.org/abs/1108.5909}
  {\path{arXiv:1108.5909}}

\bibitem{CKS00}
 K.~G. Chetyrkin, J.~H. Kuhn, M.~Steinhauser,
  \textit{RunDec: A Mathematica package for running and decoupling
  of the strong coupling and quark masses},
  Comput. Phys. Commun. 133 (2000) 43--65.
  \newblock \href {http://arxiv.org/abs/hep-ph/0004189}
  {\path{arXiv:hep-ph/0004189}}
  \href{http://dx.doi.org/10.1016/S0010-4655(00)00155-7}
  {\path{doi:10.1016/S0010-4655(00)00155-7}}

\end{thebibliography}

\newcommand{\noopsort}[1]{} \newcommand{\printfirst}[2]{#1}
 \newcommand{\singleletter}[1]{#1} \newcommand{\switchargs}[2]{#2#1}

\end{document}